\documentclass[aps,prd,nofootinbib,showpacs,preprintnumbers]{revtex4-1}
\newcommand{\be}{\begin{eqnarray}}
\newcommand{\ee}{\end{eqnarray}}

\newcommand{\hchi}{\hat{\chi}}
\def\lsim{\raise0.3ex\hbox{$<$\kern-0.75em\raise-1.1ex\hbox{$\sim$}}}
\def\gsim{\raise0.3ex\hbox{$>$\kern-0.75em\raise-1.1ex\hbox{$\sim$}}}
\usepackage{color}
\usepackage{epsfig}
\usepackage{graphics}
\usepackage{amssymb}

\newcommand{\fk}[1]{{\textcolor{black}{#1}}} 

\begin{document}

\title{Fluctuations and Correlations of net baryon number, electric
charge, and strangeness: A comparison of lattice QCD results with
the hadron resonance gas model
   }
  \author{
A. Bazavov$^{\rm a}$, 
Tanmoy Bhattacharya$^{\rm b}$, 
C. E. DeTar$^{\rm c}$, 
H.-T. Ding$^{\rm a}$, 
Steven Gottlieb$^{\rm d}$, 
Rajan Gupta$^{\rm b}$,\\
P. Hegde$^{\rm a}$, 
Urs M. Heller$^{\rm e}$, 
F. Karsch$^{\rm a,f}$,
E. Laermann$^{\rm f}$, 
L. Levkova$^{\rm c}$, 
Swagato Mukherjee$^{\rm a}$, \\
P. Petreczky$^{\rm a}$, 
Christian Schmidt$^{\rm f}$, 
R.A. Soltz$^{\rm g}$, 
W. Soeldner$^{\rm h}$,
R. Sugar$^{\rm i}$, 
and Pavlos M. Vranas$^{\rm g}$
\\[1mm]
{\bf (HotQCD Collaboration)}
}
\address{
$^{\rm a}$ Physics Department, Brookhaven National Laboratory, Upton, NY 11973, USA \\
$^{\rm b}$ Theoretical Division, Los Alamos National Laboratory, Los Alamos, NM 87545, USA\\
$^{\rm c}$ Department of Physics and Astronomy, University of Utah, Salt Lake City, UT 84112, USA \\
$^{\rm d}$ Physics Department, Indiana University, Bloomington, IN 47405, USA\\
$^{\rm e}$ American Physical Society, One Research Road, Ridge, NY 11961, USA\\
$^{\rm f}$ Fakult\"at f\"ur Physik, Universit\"at Bielefeld, D-33615 Bielefeld, Germany\\
$^{\rm g}$ Physics Division, Lawrence Livermore National Laboratory, Livermore CA 94550, USA\\
$^{\rm h}$ Institut f\"ur Theoretische Physik, Universit\"at Regensburg, D-93040 Regensburg, Germany\\
$^{\rm i}$ Physics Department, University of California, Santa Barbara, CA 93106, USA\\
}
\date{\today}
\begin{abstract}

We calculate the quadratic fluctuations of net baryon number, electric
charge and strangeness as well as correlations among these conserved
charges in (2+1)-flavor lattice QCD at zero chemical potential.
Results are obtained using calculations with
tree level improved gauge and the highly improved staggered quark
(HISQ) actions with almost physical light and strange quark masses at
three different values of the lattice cut-off. 
Our choice of parameters corresponds to a value of $160$~MeV for
the lightest pseudo scalar Goldstone mass and a physical value of 
the kaon mass.
The three diagonal charge susceptibilities 
\fk{and the correlations among conserved charges}
have been extrapolated to the continuum limit in the temperature 
interval $150~{\rm MeV} \le T \le 250~{\rm MeV}$. 
We compare our results
with the hadron resonance gas (HRG) model calculations and find
agreement with HRG model results only for temperatures
$T\ \lsim\ 150$~MeV. We observe significant deviations in the
temperature range $160~{\rm MeV}\ \lsim\ T\ \lsim\ 170~{\rm MeV}$ and
qualitative differences in the behavior of the three conserved charge
sectors.  At $T\simeq 160$~MeV quadratic net baryon number
fluctuations in QCD agree with HRG model calculations while, the net
electric charge fluctuations in QCD are about 10\% smaller and net
strangeness fluctuations are about 20\% larger.  These findings are 
relevant to the discussion of freeze-out conditions in relativistic
heavy ion collisions.

\end{abstract}

\maketitle

\section{Introduction}

The low energy runs currently being performed at the Relativistic
Heavy Ion Collider (RHIC) \cite{RHIC} aim at an exploration of the QCD
phase diagram at non-zero temperature ($T$) and baryon chemical
potential ($\mu_B$) through the measurement of fluctuations of
conserved charges, $e.g.$, net baryon number, electric charge and
strangeness. For the former, first results have been published by the
STAR collaboration \cite{STAR} and preliminary results on the latter
two have been presented at conferences. A central goal of these
experiments is to search for the existence of the QCD critical point,
a second order phase transition point, that has been postulated to
exist at non-vanishing baryon chemical potential in the $T$-$\mu_B$
phase diagram of QCD \cite{CEP,CP}. Fluctuations in conserved charges can
probe this critical point, the endpoint of a line of first order phase
transitions that extends to large baryon chemical potential at
non-zero quark masses.  More generally, the study of
fluctuations at any value of the baryon chemical potential 
probe thermal conditions in a medium and provide information on the 
critical behavior of QCD~\cite{Karsch:2011}.

\fk{
At zero baryon chemical potential, already, the
analysis of fluctuations of conserved charges and their higher order
cumulants provides important information about the relation between
the QCD chiral phase transition at vanishing light quark
masses, the cross-over temperature at physical quark masses and
the freeze-out conditions observed in heavy ion experiments
\cite{Karsch:2011,Redlich,Friman}.
Thus, calculations of conserved charge
fluctuations at $\mu_B=0$ will provide unique information on
freeze-out conditions at the LHC where the baryon chemical potential
is small, $\mu_B/T \approx 0.05$.  The quadratic fluctuations of the net
baryon number characterize the width of the probability distribution
which has been measured at RHIC
\cite{STAR}
and has recently been
analyzed in the framework of the hadron resonance gas (HRG) model
\cite{PBM}.
Although the HRG model provides a rather
satisfactory description of global hadron yields at chemical
freeze-out
\cite{HRG},
its ability to describe detailed properties of strongly
interacting matter such as fluctuations of conserved charges and, in
particular, their higher order cumulants is not obvious.
The foundation for the HRG model is given by the Dashen, Ma, Bernstein
theorem
\cite{Dashen},
which shows that the partition function of
strongly interacting matter can be described by a gas of free
resonances, if the system is sufficiently dilute and
the resonance production is the dominant part of the interaction among
hadrons \cite{Raju,Becattini}.
At very low temperature and high baryon number density,
where nonresonant nucleon-nucleon interactions become important,
as well as at high temperature where strongly interacting matter
undergoes a transition to the quark-gluon plasma regime and partonic
degrees of freedom become dominant, the HRG
model is expected to be a poor approximation to the thermodynamics
of strongly interacting matter. To what extent the HRG model 
provides a good description of strongly interacting matter needs to 
be explored in detail
by comparing model calculations with first principal (lattice) QCD calculations.
The latter provides the complete description of
QCD thermodynamics at all values of the temperature and ultimately
should set the standard for the interpretation of experimental
results on strong interaction thermodynamics. Unfortunately, this
is, at present, not fully  possible. For instance, the only
satisfactory way to specify the thermal conditions in a heavy
ion experiment at the time of hadronization is through the comparison
of experimental data with HRG model calculations
\cite{Cleymans}.
It has recently been pointed out that in the future this may be
overcome by comparing experimental data on conserved charge
fluctuations directly with lattice QCD calculations 
\cite{Karsch:2011}.
Also in order to establish such an approach more firmly it is
important to understand and quantify to what extent lattice
QCD calculations and HRG model calculations agree and in which
temperature regime the latter provides a reasonable approximation
to strong interaction thermodynamics.
}

\fk{
Indeed, some
deviations in ratios of higher order cumulants of baryon number
fluctuations, calculated within the HRG model, from experimental
results for cumulants of net proton fluctuations have been observed
\cite{PBM}. Whether these deviations can be accounted for within QCD
or are of more technical origin related to the restricted phase space
in which experimental observations have been performed is an open
question. In any case, a more detailed analysis of the thermal
conditions achieved in heavy ion experiments is important. Lattice QCD
calculations of fluctuations of conserved charges in equilibrium
thermodynamics provide a base line for such discussions.  Studies
of fluctuations and higher order cumulants \cite{Ejiri} may reveal
differences between HRG model calculations and QCD thermodynamics that
will appear close to criticality in the QCD phase
diagram.
}

\fk{
In this paper we will use the HRG model in its simplest version,
i.e. as a sum of non-interacting, point-like particles.
This is known to provide an accurate description of a dilute, strongly
interacting hadron gas \cite{Raju}.
The need for taking into account residual
interactions, for instance through the introduction of an intrinsic
size of the hadrons \cite{Andronic} has been discussed. It has also
been noted that the inclusion of higher mass resonances and an
improvement in the strangeness sector of the HRG model may be needed
to adequately describe pronounced features of hadron production such
as the enhancement in the $K/\pi$ ratio \cite{NewHRG}.
The advantage of the simplest version, however, is that the HRG model
in this form is parameter free, while any improvement on this model
will introduce further parameters without bringing us closer to the
actual underlying theory, QCD, in a controlled way.
}

Quadratic fluctuations of conserved charges are closely related to
quark number susceptibilities \cite{Gottlieb}.  Fluctuations of net
baryon number, electric charge and strangeness, as well as correlations
among them, have been analyzed in previous lattice QCD calculations
\cite{Gavai,Bernard,RBCBI,Fodor_11} and have also been used to
characterize properties of the relevant thermodynamic degrees of
freedom at low as well as high temperature \cite{Ejiri,Koch1,Allton}.
The generic forms of their temperature dependence and their scaling
properties are understood in terms of universal properties of the QCD
partition function and its derivatives in the vicinity of the QCD
chiral phase transition \cite{Friman,RBCBI}. To make use of this
knowledge in a quantitative comparison with experimental results,
lattice QCD calculations close to the continuum are needed.

In this paper we present an analysis of fluctuations in, 
and correlations among, conserved charges using numerical calculations
in (2+1)-flavor QCD at three values of the lattice
cut-off~\footnote{Preliminary results of this work had been presented
at Quark Matter 2011 \cite{Mukherjee} and PANIC 2011 \cite{Hegde}.}.
For these calculations we exploit an ${\cal O}(a^2)$ improved action
consisting of a tree-level improved gauge action combined with the
highly improved staggered fermion action (HISQ/tree)
\cite{Follana,hotQCDTc}.  We discuss the cut-off dependence of our
results in different temperature intervals and consider two different
zero-temperature observables for the determination of the temperature
scale used for extrapolations to the continuum limit.  This allows us
to quantify systematic errors in our calculation.  In an
appendix, we discuss the relation between temperature scales deduced
from different zero-temperature observables and the propagation of
their cut-off dependence into the cut-off dependence of thermodynamic
observables.

\section{Fluctuations of conserved charges from lattice QCD; the 
hadron resonance gas and the ideal gas limit}

To calculate fluctuations of baryon number ($B$), electric charge ($Q$) and 
strangeness ($S$) from (lattice) QCD we start from the
QCD partition function with non-zero light ($\mu_u$, $\mu_d$) and strange 
quark ($\mu_{s}$) chemical potentials.  The quark chemical
potentials can be expressed in terms of chemical potentials for
baryon number ($\mu_B$), strangeness ($\mu_S$) and electric charge ($\mu_Q$),
\begin{eqnarray}
\mu_u&=&\frac{1}{3}\mu_B + \frac{2}{3}\mu_Q \; , \nonumber \\
\mu_d&=&\frac{1}{3}\mu_B - \frac{1}{3}\mu_Q \; ,\nonumber \\
\mu_s&=&\frac{1}{3}\mu_B - \frac{1}{3}\mu_Q - \mu_S \; .
\label{potential}
\end{eqnarray}
The starting point of the analysis is the pressure $p$ given by the logarithm of the QCD partition function,
\begin{equation}
\frac{p}{T^4} \equiv \frac{1}{VT^3}\ln Z(V,T,\mu_B,\mu_S,\mu_Q) \; .
\label{pressure}
\end{equation}
Fluctuations of the conserved charges and their correlations in a thermalized
medium are then obtained
from its derivatives evaluated at $\vec{\mu}=(\mu_{B},\mu_{Q},\mu_{S})=0$,
\begin{eqnarray}
\hat{\chi}_2^X \equiv \frac{\chi_2^X}{T^2} &=& \left. \frac{\partial^2p/T^4}{\partial\hat{\mu}_X^2}  
\right|_{\vec\mu =0} \;\; , 
\label{obsX2}
\\
\hat{\chi}_{11}^{XY} \equiv \frac{\chi_{11}^{XY}}{T^2} &=& \left. 
\frac{\partial^{2}p/T^4}{\partial\hat{\mu}_X\partial\hat{\mu}_Y}\right|_{\vec\mu =0} \; ,
\label{obs}
\end{eqnarray}
with $\hat\mu_X \equiv \mu_X/T$ and $X,\ Y= B,\ Q,\ S$. 
Explicit expressions for the calculation of these susceptibilities in terms
of generalized light and strange quark number susceptibilities are given
in \cite{RBCBI}.   

As all these derivatives
are evaluated at $\vec\mu =0$, the expectation values of all net charge numbers
$\delta N_X\equiv N_X-N_{\bar{X}}$, with $N_X$ ($N_{\bar{X}}$), denoting
the number of particles (anti-particles), vanish, {\it i.e.,}
$\langle \delta N_X \rangle =0$.
The susceptibilities, {\it i.e.}, the quadratic
fluctuations of the charges, are then given by 
\begin{equation}
\hat{\chi}_2^X = \langle (\delta N_X)^2 \rangle/VT^3 \ .
\end{equation}

\subsection{The hadron resonance gas}
We will compare results for fluctuations and correlations defined by 
Eqs.~(\ref{obsX2}) and (\ref{obs}) with hadron resonance gas model calculations.
The partition function of the HRG model can be split into mesonic and
baryonic contributions,
\begin{eqnarray}
\frac{p^{HRG}}{T^4} \hspace{-2mm}
&=&\frac{1}{VT^3}\sum_{i\in\;mesons}\hspace{-3mm} 
\ln{\cal Z}^{M}_{M_i}(T,V,\mu_Q,\mu_S)
\nonumber \\
&&\hspace{-3mm} +\frac{1}{VT^3}
\sum_{i\in\;baryons}\hspace{-3mm} \ln{\cal Z}^{B}_{M_i}(T,V,\mu_B,\mu_Q,\mu_S)\; ,
\label{ZHRG}
\end{eqnarray}
where the partition function for mesonic ($M$) or baryonic ($B$) particle 
species $i$ with mass $M_i$ is given by,
\begin{eqnarray}
\ln{\cal Z}^{M/B}_{M_i}
&=&\mp \frac{V d_i}{{2\pi^2}} \int_0^\infty dk k^2
\ln(1\mp z_ie^{-\varepsilon_i/T}) \nonumber \; \\
&=& 
{{VT^3}\over
{2\pi^2}}d_i \left(\frac{M_i}{T}\right)^2
\sum_{k=1}^\infty (\pm 1)^{k+1} {z_i^k \over {k^2}} K_2({{k M_i/T}}) 
\quad .
\label{ZMB}
\end{eqnarray}
Here upper signs correspond to mesons and lower signs to baryons;
$\varepsilon_i=\sqrt{k^2+M_i^2}$ denotes the energy of particle $i$, 
$d_i$ is its degeneracy factor and its fugacity is given by
\begin{equation}
z_i=\exp\left((B_i\mu_B+Q_i\mu_Q+S_i\mu_S)/T\right) \; .
\label{fuga}
\end{equation}
With these relations it is straightforward to calculate susceptibilities and 
charge correlations in the HRG model using Eqs.~(\ref{obsX2}) and (\ref{obs}).

We note that a HRG model is defined by specifying the resonance
spectrum used to construct the partition function in 
Eq.~(\ref{ZHRG}). 
We use all hadron resonances with masses $M_H\le 2.5$~GeV
listed by the particle data group (PDG) in their 2010 summary 
tables\footnote{In this summary table there are a few
three starred resonances listed which do not have a known spin assignment.
For these we use the minimal degeneracy factors. We also checked that the inclusion
of some known heavier resonances as well as the inclusion of charmed hadrons
does not alter the picture presented here.
\fk{Moreover, we checked the stability of HRG results by reducing
the mass cut-off from 2.5~GeV to 2.0~GeV. This alters the relevant
observables discussed here by at most 2\% at $T=200$~MeV.}
} \cite{PDG}.
It is similar to that used, for example, in Ref.~\cite{NewHRG}. 
One question, that we will discuss in the
comparison of lattice QCD results with HRG model calculations, 
is to what extent the strangeness
sector is well represented in the HRG model calculations. 
This question has also been addressed in \cite{NewHRG}.

\subsection{The ideal gas limit}

In the infinite temperature limit, the grand canonical QCD partition
function reduces to that of an ideal gas of quarks and gluons.
In this limit, quark mass effects, including those in the strange
quark sector, are negligible and we may compare our (2+1)-flavor QCD calculations with
a free quark-gluon gas of 3-flavor QCD (Stefan-Boltzmann (SB) gas). 
This is given by \cite{Kapusta}

\begin{eqnarray}
\frac{p_{SB}}{T^4} 
=  \frac{8 \pi^2}{45} + 
\frac{7 \pi^2}{20} +
 \sum_{f=u,d,s} \left[
\frac{1}{2}  \left(\frac{\mu_f}{T}\right)^2 
+ \frac{1}{4 \pi^2} \left(\frac{\mu_f}{T}\right)^4 
\right] \quad ,
\label{eq:free}
\end{eqnarray}
where the first two terms give the contributions of the gluon and
the quark sectors for vanishing chemical potentials. After 
expressing the flavor chemical potentials in terms of $\mu_B$, $\mu_Q$
and $\mu_S$ as given in Eq.~(\ref{potential}), it is straightforward to 
read off the ideal gas values for diagonal and off-diagonal susceptibilities.
These are listed in Table~\ref{tab:SB_coefficients}.

\begin{table}[t]
\begin{center}
\begin{tabular}{|c|c|c|c|}
\hline
    & $B$ & $Q$ & $S$ \\
\hline
$B$ & 1/3 &  0  &-1/3 \\
\hline
$Q$ &  0  & 2/3 & 1/3 \\
\hline
$S$ &-1/3 & 1/3 &  1  \\
\hline
\end{tabular}
\end{center}
\caption{Ideal gas values for off-diagonal,
$\hchi_{11}^{XY}\equiv \chi_{11}^{XY}/T^2$, and diagonal susceptibilities, 
$\hchi_2^X\equiv \hchi_{11}^{XX}$. Here $X,\ Y=B,\ Q,\ S$.}
\label{tab:SB_coefficients}
\end{table}

\section{Lattice calculations}
\label{sec:lattice}

In order to analyze fluctuations of conserved charges, we perform
calculations on gauge field configurations generated in our study of
the finite-temperature transition in (2+1)-flavor QCD \cite{hotQCDTc}.
These calculations were performed with the HISQ/tree action for three
values of the lattice cut-off corresponding to lattices with temporal
extent $N_\tau=6,\ 8$ and $12$ and a spatial lattice extent $N_\sigma
= 4 N_\tau$.  The calculations cover a temperature range, $130~{\rm
  MeV}\ \lsim\ T\ \lsim \ 350~{\rm MeV}$.  In this temperature range
the line of constant physics is defined by tuning the strange quark
mass $m_s$ to its physical value and setting the light quark masses
to $m_l= m_s/20$, which correspond to a pion mass $M_\pi \simeq
160$~MeV.  

It is important to note that the definition of physical quark and/or
pion masses is not straightforward in calculations with staggered
fermions at non-zero values of the lattice spacing due to taste
symmetry breaking. Taste symmetry breaking, which is a consequence of
the doubling problem in the staggered formulation, gives rise to
sixteen pseudo-scalar mesons corresponding to the sixteen elements of
the Clifford algebra, of which only one, with taste matrix $\gamma_5$,
behaves as a Goldstone particle at finite lattice spacing.  Even in
the chiral limit, the other 15 modes receive masses of ${\cal O}(a^2)$
which vanish only in the continuum limit.  Thus, as a result of taste
symmetry breaking, these sixteen modes contribute to observables with
different masses. The same is true of all
other states, but the problem is most severe for Goldstone modes. The
size of these effects has been quantified for the staggered fermion
discretization scheme exploited here (HISQ/tree)
by measuring the masses of
the sixteen taste pions and by defining a root-mean-squared
mass~\cite{hotQCDTc},
\begin{equation}
  M_\pi^{RMS}= \frac{1}{4}\sqrt{
  M_{\gamma_5}^2+M_{\gamma_0\gamma_5}^2
  +3M_{\gamma_i\gamma_5}^2+3M_{\gamma_i\gamma_j}^2
  +3M_{\gamma_i\gamma_0}^2+3M_{\gamma_i}^2
  +M_{\gamma_0}^2+M_{1}^2} \, .
\label{piRMS}
\end{equation}
Taste symmetry breaking also affects the generation of the ensemble of gauge
configurations on which measurements are made. To simulate the desired
number of flavors, the fourth-root of the staggered fermion
determinant is taken for each flavor. While this ``rooting'' trick
corrects for the number of flavors in the continuum limit, at finite
lattice spacings, the masses of states contributing to the partition
function are not degenerate.

With the improved staggered fermion action (HISQ/tree) used in our
calculation, taste violations, while strongly
suppressed, are still large~\cite{hotQCDTc}. In fact, our analysis showed that the
HISQ/tree action has the smallest taste violations compared to the
other improved staggered actions (p4, asqtad and stout) used in finite
temperature calculations.  Nevertheless, at values of the cut-off
corresponding to the transition region ($T\simeq 160$~MeV) on the
three different $N_\tau$ lattices analyzed by us, the RMS masses vary
from $M_\pi^{RMS}\simeq 215$~MeV on our finest lattices ($N_\tau =12$)
to $M_\pi^{RMS}\simeq 415$~MeV on the coarsest lattice ($N_\tau =6$),
in contrast to the Goldstone pion mass $140$ MeV~\cite{hotQCDTc}. We also find
that, to a good approximation, the difference of $M_\pi^{RMS}$ and the
physical pion mass $M_\pi$ is proportional to the square of the
lattice cut-off at a fixed value of the temperature, $(aT)^2 \sim
1/N_\tau^2$.  In this paper we show that the cut-off effects resulting
from a much heavier RMS mass influence most strongly the electric
charge fluctuations as these are most sensitive to the 
contributions from pions.  It is worth pointing out, for comparison, 
that to obtain an RMS mass of about $215$~MeV achieved with the
HISQ/tree action on $N_\tau=12$ lattices will require $N_\tau \approx
20$ with the asqtad and stout actions~\cite{hotQCDTc,Fodor_11}.

We typically analyzed 10,000--20,000 gauge field configurations per
parameter set for $N_\tau=6$, 10,000--14,000 configurations
for $N_\tau=8$ and up to 6000 configurations for
$N_\tau=12$ lattices. Measurements of all operators needed to calculate
quadratic fluctuations have been performed every 10 hybrid Monte Carlo
time units for $N_\tau=12$ and every 10 or 20 time units for $N_\tau =
6,\ 8$ lattices.  Most of our quark number susceptibilities on the $N_\tau=6$
and $8$ lattices were calculated on GPU-clusters and we used
500--1500 random source vectors for the analysis. 
The $N_\tau=12$ data were analyzed using 400 random source
vectors at low temperatures and 100 at high temperatures.

\begin{figure}[t]
\begin{center}
\epsfig{file=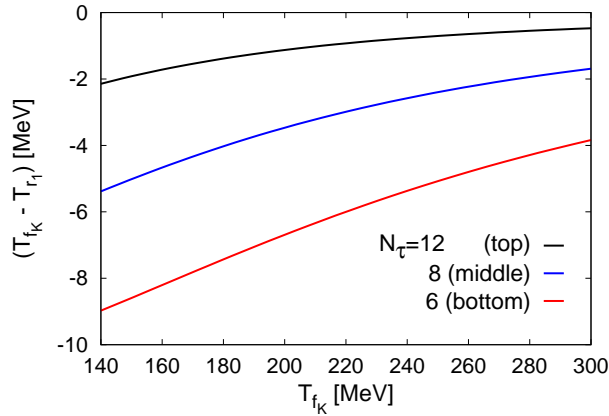,width=8.5cm}
\caption{Difference in temperature scales obtained from calculations of
$r_1$ and $f_K$ at values of the cut-off relevant for the temperature
range explored in the calculation of quark number susceptibilities.
}
\label{fig:cutoff}
\end{center}
\end{figure}

When performing lattice QCD calculations at non-zero temperature, we
have to control (at least) two different sources of cut-off errors. On
the one hand there is the intrinsic cut-off dependence of the
observables calculated at non-zero temperature at a certain value of
the cut-off $a^{-1}$. We reduce these by working with tree-level
${\cal O} (a^2)$ improved actions in the gauge as well as the fermion
sector. This improvement also insures that at high temperature the
cut-off dependence of the basic operators entering our calculations is
small.  An additional cut-off dependence arises due to the choice of
the zero temperature observable used to set the scale for all finite
temperature measurements. We investigate two different
scale-setting observables: the length scale $r_1$ extracted from
the slope of the static quark potential,
\begin{equation}
\left( r^2 \frac{{\rm d}V_{\bar{q}q}(r)}{{\rm d}r} \right)_{r=r_1} =
1.0  \;\; ,
\label{r0r1}
\end{equation}
and the kaon decay constant $f_K$. The temperature in these units is
$T_{r_1} r_1= r_1/(a N_\tau)$ and $T_{f_K}/f_K = 1/(f_K a N_\tau)$.
The efficacy of these observables in setting the scale is
complementary in many respects~\cite{hotQCDTc}. In particular, we note
that $r_1$ has a mild dependence on the quark masses and is well
defined even in the infinite quark mass limit. The kaon decay
constant, on the other hand, has a sizeable quark mass dependence, and
at lattice spacings used in this study, it is sensitive to taste
symmetry violations in the hadron spectrum.  To convert to physical
units, we use $r_1= 0.3106(20)$~fm \cite{milcr1} and the latest PDG
value for the kaon decay constant, $f_K=156.1/\sqrt{2}$~MeV.  In
Fig.~\ref{fig:cutoff}, we show the difference in the two estimates of
temperature over the interval relevant to the calculations performed
here. The leading order correction contributing to this difference is
${\cal O}(g^2 a^2)$.

As noted in~\cite{hotQCDTc}, we also find that in the study of charge
fluctuations it is advantageous to use the $f_K$ scale as it
absorbs a significant fraction of the cut-off effects, $i.e.$, the
cut-off effects are similar and cancel to a large extent in the ratio of 
hadron masses and temperature, $M/T$.  However, it
should be stressed that any one observable ($r_1$ or $f_K$) cannot
eliminate cut-off effects in all observables equally well. We
elaborate on this point in more detail in the appendix.

\section{Fluctuations}

\subsection{Fluctuations in baryon number, strangeness, and electric charge}
\label{ssec:fluctuations}

We start our discussion of the fluctuations of baryon number,
strangeness and electric charge by summarizing the data obtained on
lattices with temporal extent $N_\tau=6,\ 8$ and $12$ in
Tables~\ref{tab:chi2_6}, \ref{tab:chi2_8} and \ref{tab:chi2_12} and
discussing their scaling behavior.  The continuum extrapolation and
comparison to the HRG model as well as the asymptotic high temperature
ideal gas results are discussed in detail in the next subsection.

 \begin{table}
 \begin{tabular}{|cc|ccc|ccc|}
 \hline
 $\beta$ & $T$ (MeV)& $\chi_2^B/T^2$                  & $\chi_2^Q/T^2$ & $\chi_2^S/T^2$                                 & $\chi_{11}^{BS}/T^2$ & $\chi_{11}^{BQ}/T^2$ &                    $\chi_{11}^{QS}/T^2$                                              \\
 \hline
 5.900 & 124.17 &   0.0223(15)  &   0.1173( 7)  &   0.0601( 6)  &  -0.0090( 6)  &   0.0067( 6)  &   0.0256( 2) \\
 6.000 & 138.22 &   0.0436(28)  &   0.1829(12)  &   0.1080(10)  &  -0.0207(11)  &   0.0114( 9)  &   0.0437( 3) \\
 6.025 & 141.98 &   0.0545(22)  &   0.2037(16)  &   0.1241(14)  &  -0.0258(12)  &   0.0144( 6)  &   0.0491( 6) \\
 6.050 & 145.83 &   0.0590(23)  &   0.2283( 7)  &   0.1426(11)  &  -0.0283(14)  &   0.0154( 6)  &   0.0572( 4) \\
 6.075 & 149.79 &   0.0745(26)  &   0.2551(14)  &   0.1672(14)  &  -0.0382(13)  &   0.0182( 6)  &   0.0645( 5) \\
 6.100 & 153.85 &   0.0881(24)  &   0.2885(14)  &   0.1983(14)  &  -0.0472(13)  &   0.0205( 6)  &   0.0756( 4) \\
 6.125 & 158.01 &   0.1044(19)  &   0.3216(18)  &   0.2317(19)  &  -0.0580(11)  &   0.0225( 4)  &   0.0868( 5) \\
 6.150 & 162.28 &   0.1256(22)  &   0.3614(31)  &   0.2741(26)  &  -0.0729(15)  &   0.0264( 4)  &   0.1006( 9) \\
 6.175 & 166.66 &   0.1466(52)  &   0.3956(32)  &   0.3200(54)  &  -0.0887(29)  &   0.0290(25)  &   0.1157(20) \\
 6.195 & 170.25 &   0.1548(32)  &   0.4176(20)  &   0.3513(30)  &  -0.0987(23)  &   0.0280( 5)  &   0.1263( 6) \\
 6.215 & 173.90 &   0.1730(34)  &   0.4427(25)  &   0.3937(38)  &  -0.1155(25)  &   0.0288( 5)  &   0.1391( 7) \\
 6.245 & 179.52 &   0.1881(29)  &   0.4666(23)  &   0.4422(36)  &  -0.1324(20)  &   0.0278( 5)  &   0.1549(10) \\
 6.285 & 187.27 &   0.2048(68)  &   0.4978(30)  &   0.5116(64)  &  -0.1504(50)  &   0.0272(32)  &   0.1806(31) \\
 6.341 & 198.61 &   0.2272(20)  &   0.5198(24)  &   0.5893(40)  &  -0.1845(17)  &   0.0214( 2)  &   0.2024(13) \\
 6.354 & 201.34 &   0.2326(53)  &   0.5261(25)  &   0.6098(72)  &  -0.1936(46)  &   0.0195(24)  &   0.2083(30) \\
 6.423 & 216.33 &   0.2517(14)  &   0.5466(12)  &   0.6881(23)  &  -0.2212(13)  &   0.0152( 1)  &   0.2334( 7) \\
 6.488 & 231.33 &   0.2639(28)  &   0.5586(15)  &   0.7420(43)  &  -0.2404(26)  &   0.0118(13)  &   0.2508(18) \\
 6.515 & 237.81 &   0.2676( 9)  &   0.5629( 9)  &   0.7576(17)  &  -0.2472( 9)  &   0.0102( 1)  &   0.2552( 5) \\
 6.550 & 246.45 &   0.2672(41)  &   0.5678(19)  &   0.7787(54)  &  -0.2528(36)  &   0.0072(19)  &   0.2629(24) \\
 6.664 & 276.43 &   0.2777(11)  &   0.5714(11)  &   0.8101(23)  &  -0.2667(11)  &   0.0055( 1)  &   0.2717( 5) \\
 6.800 & 316.10 &   0.2806(18)  &   0.5718(10)  &   0.8248(36)  &  -0.2722(18)  &   0.0042( 8)  &   0.2761(14) \\
 6.950 & 365.18 &   0.2862(27)  &   0.5714(15)  &   0.8374(57)  &  -0.2794(26)  &   0.0034(14)  &   0.2790(21) \\
 7.150 & 440.31 &   0.2803(23)  &   0.5666(12)  &   0.8395(43)  &  -0.2783(23)  &   0.0010(11)  &   0.2804(17) \\
 \hline
 \end{tabular}

 \caption{Quadratic fluctuations of net baryon number, electric charge and 
strangeness as well as correlations among these conserved charges
in units of $T^2$ calculated on lattices with temporal extent $N_\tau=6$.
We use $f_K$ to define the temperature scale.}
 \label{tab:chi2_6}
 \end{table}

\begin{table}
 \begin{tabular}{|cc|ccc|ccc|}
 \hline
 $\beta$ & $T$ (MeV)& $\chi_2^B/T^2$                  & $\chi_2^Q/T^2$ & $\chi_2^S/T^2$                                 & $\chi_{11}^{BS}/T^2$ & $\chi_{11}^{BQ}/T^2$ &                    $\chi_{11}^{QS}/T^2$                                              \\
 \hline
 6.195 & 127.69 &   0.0083(58)  &   0.1377(19)  &   0.0690(13)  &  -0.0050(21)  &   0.0018(20)  &   0.0321( 7) \\
 6.245 & 134.64 &   0.0332(52)  &   0.1711(18)  &   0.0933(17)  &  -0.0151(23)  &   0.0090(16)  &   0.0391( 7) \\
 6.260 & 136.79 &   0.0390(28)  &   0.1868(20)  &   0.1054(11)  &  -0.0183(10)  &   0.0104(12)  &   0.0436( 4) \\
 6.285 & 140.45 &   0.0417(40)  &   0.2092(19)  &   0.1212(18)  &  -0.0198(20)  &   0.0110(11)  &   0.0507( 3) \\
 6.315 & 144.95 &   0.0668(38)  &   0.2455(23)  &   0.1490(31)  &  -0.0315(29)  &   0.0177( 7)  &   0.0588( 6) \\
 6.341 & 148.96 &   0.0673(49)  &   0.2692(38)  &   0.1718(39)  &  -0.0345(29)  &   0.0164(12)  &   0.0686( 7) \\
 6.354 & 151.00 &   0.0705(63)  &   0.2908(23)  &   0.1878(31)  &  -0.0359(35)  &   0.0173(15)  &   0.0760( 7) \\
 6.390 & 156.78 &   0.0988(54)  &   0.3362(37)  &   0.2385(40)  &  -0.0557(31)  &   0.0215(11)  &   0.0914(10) \\
 6.423 & 162.25 &   0.1235(28)  &   0.3811(15)  &   0.2894(23)  &  -0.0731(18)  &   0.0252( 7)  &   0.1081( 6) \\
 6.445 & 165.98 &   0.1444(27)  &   0.4161(30)  &   0.3398(19)  &  -0.0919(15)  &   0.0263( 8)  &   0.1239(11) \\
 6.460 & 168.57 &   0.1610(24)  &   0.4287(19)  &   0.3623(30)  &  -0.1034(17)  &   0.0288( 5)  &   0.1294( 9) \\
 6.488 & 173.49 &   0.1731(34)  &   0.4534(24)  &   0.4058(44)  &  -0.1170(28)  &   0.0281( 4)  &   0.1444(10) \\
 6.515 & 178.36 &   0.1948(24)  &   0.4794(28)  &   0.4606(35)  &  -0.1390(19)  &   0.0279( 5)  &   0.1608(11) \\
 6.550 & 184.84 &   0.2085(17)  &   0.5034(19)  &   0.5177(30)  &  -0.1572(13)  &   0.0257( 4)  &   0.1803(10) \\
 6.575 & 189.58 &   0.2190(17)  &   0.5171(10)  &   0.5586(19)  &  -0.1719(13)  &   0.0236( 2)  &   0.1933( 4) \\
 6.608 & 196.01 &   0.2291(23)  &   0.5321(18)  &   0.6052(39)  &  -0.1874(22)  &   0.0209( 1)  &   0.2089(10) \\
 6.664 & 207.32 &   0.2465(22)  &   0.5511(13)  &   0.6711(35)  &  -0.2123(20)  &   0.0171( 2)  &   0.2294( 8) \\
 6.800 & 237.07 &   0.2748(14)  &   0.5786(14)  &   0.7796(27)  &  -0.2540(13)  &   0.0104( 1)  &   0.2628( 7) \\
 6.950 & 273.88 &   0.2901( 7)  &   0.5963(10)  &   0.8443(18)  &  -0.2782( 7)  &   0.0060( 1)  &   0.2831( 5) \\
 7.150 & 330.23 &   0.2984( 5)  &   0.6073( 9)  &   0.8829(13)  &  -0.2919( 5)  &   0.0032( 1)  &   0.2955( 4) \\
 \hline
 \end{tabular}

 \caption{Same as Table~\protect\ref{tab:chi2_6} but for $N_\tau=8$.}
 \label{tab:chi2_8}
 \end{table}

 \begin{table}
 \begin{tabular}{|cc|ccc|ccc|}
 \hline
 $\beta$ & $T$ (MeV)& $\chi_2^B/T^2$                  & $\chi_2^Q/T^2$ & $\chi_2^S/T^2$                                 & $\chi_{11}^{BS}/T^2$ & $\chi_{11}^{BQ}/T^2$ &                    $\chi_{11}^{QS}/T^2$                                              \\
 \hline
  6.70 & 143.25 &   0.0404(135)  &   0.2741( 64)  &   0.1586( 53)  &  -0.0227(59)  &   0.0088(65)  &   0.0679(33) \\
  6.74 & 149.03 &   0.0764( 67)  &   0.3178( 52)  &   0.1939( 46)  &  -0.0348(40)  &   0.0208(32)  &   0.0796(25) \\
  6.77 & 153.48 &   0.1031( 68)  &   0.3492( 86)  &   0.2365( 59)  &  -0.0581(42)  &   0.0225(34)  &   0.0892(33) \\
  6.80 & 158.05 &   0.1056( 83)  &   0.3721( 46)  &   0.2663( 53)  &  -0.0632(47)  &   0.0212(40)  &   0.1015(27) \\
  6.84 & 164.31 &   0.1520( 42)  &   0.4358( 45)  &   0.3531( 59)  &  -0.0975(30)  &   0.0272(20)  &   0.1278(22) \\
  6.88 & 170.77 &   0.1522( 64)  &   0.4471( 40)  &   0.3808( 66)  &  -0.1009(46)  &   0.0257(30)  &   0.1399(30) \\
  6.91 & 175.76 &   0.1920( 46)  &   0.4943( 36)  &   0.4650( 63)  &  -0.1361(42)  &   0.0280(21)  &   0.1645(28) \\
  6.95 & 182.59 &   0.2034( 55)  &   0.5030( 59)  &   0.5067( 90)  &  -0.1506(55)  &   0.0264(28)  &   0.1780(46) \\
  6.99 & 189.64 &   0.2273( 73)  &   0.5273( 69)  &   0.5779( 87)  &  -0.1820(66)  &   0.0227(35)  &   0.1979(48) \\
  7.03 & 196.91 &   0.2470( 49)  &   0.5483( 36)  &   0.6319( 87)  &  -0.2006(45)  &   0.0232(24)  &   0.2157(34) \\
  7.10 & 210.21 &   0.2547( 33)  &   0.5671( 27)  &   0.7004(106)  &  -0.2213(38)  &   0.0167(16)  &   0.2395(30) \\
  7.15 & 220.15 &   0.2650( 25)  &   0.5744( 26)  &   0.7389( 44)  &  -0.2366(32)  &   0.0142(12)  &   0.2512(24) \\
  7.28 & 247.91 &   0.2822( 19)  &   0.5902( 15)  &   0.8063( 37)  &  -0.2646(22)  &   0.0088( 8)  &   0.2708(15) \\
 \hline
 \end{tabular}

 \caption{Same as Table~\protect\ref{tab:chi2_6} but for $N_\tau=12$.}
 \label{tab:chi2_12}
 \end{table}

\begin{figure}[t]
\begin{center}
\hspace*{-0.6cm}\epsfig{file=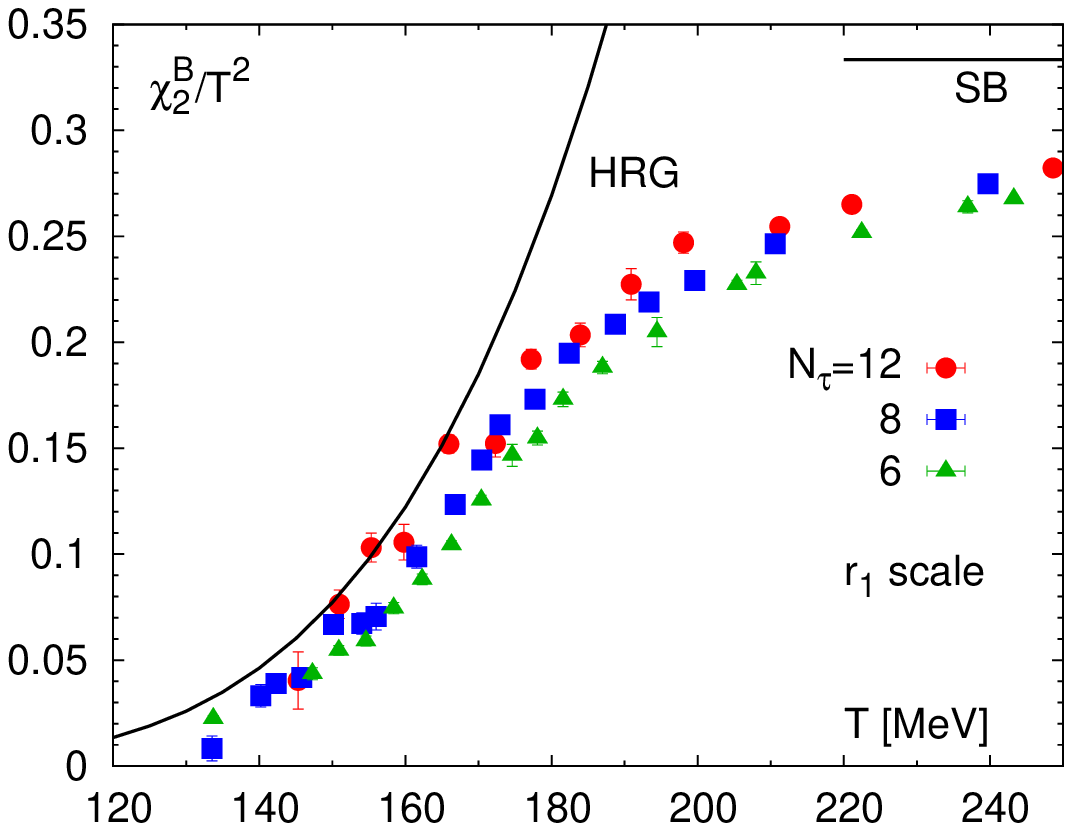,width=7.5cm}\hspace*{-0.6cm}
\epsfig{file=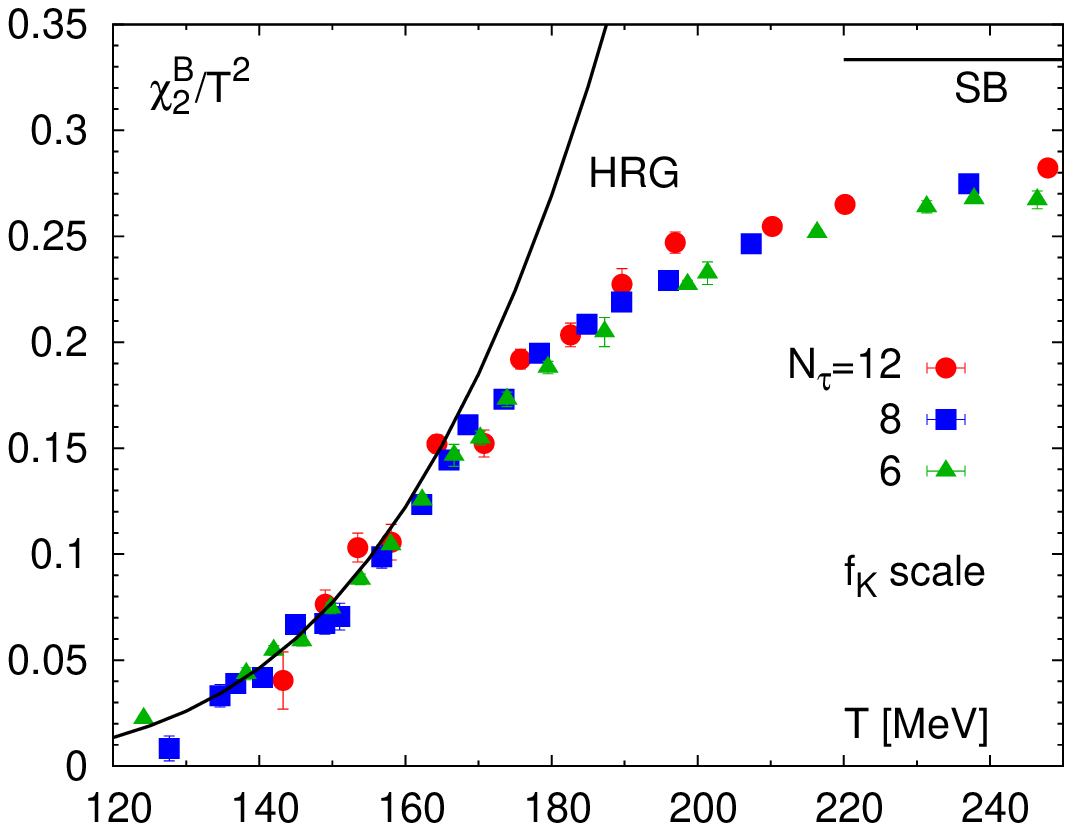,width=7.5cm}
\end{center}
\caption{
Net baryon number fluctuations versus temperature. The left hand
figure shows results using the potential shape parameter $r_1$ to set
the scale for the temperature. The right hand figure shows the same data
using $f_K$ to set the scale. Also shown are the results obtain from 
the HRG model and the infinite temperature ideal gas limit (solid lines).
}
\label{fig:chi2B}
\end{figure}

Figure~\ref{fig:chi2B} shows results for the baryon number
susceptibility with the temperature scale set using $r_1$ (left hand
panel) and $f_K$ (right hand panel).
In both cases, we show the results from the HRG model including all
resonances with mass $M_H\le 2.5$~GeV. The noticeable differences
between the left and right hand panels of Fig.~\ref{fig:chi2B} are 
due to setting of the temperature scale and we find that the 
cut-off effects are smaller when a scale based on $f_K$ is used.  As
pointed out above, this feature has also been noted in the analysis of
chiral observables~\cite{hotQCDTc,WB}. In principle, both scales should lead
to identical results in the continuum limit, however, the continuum
extrapolation is much better controlled when the cut-off effects are
small. We therefore use the $f_K$ temperature scale in the rest of
the paper unless stated otherwise. 

\begin{figure}[t]
\begin{center}
\hspace*{-0.6cm}\epsfig{file=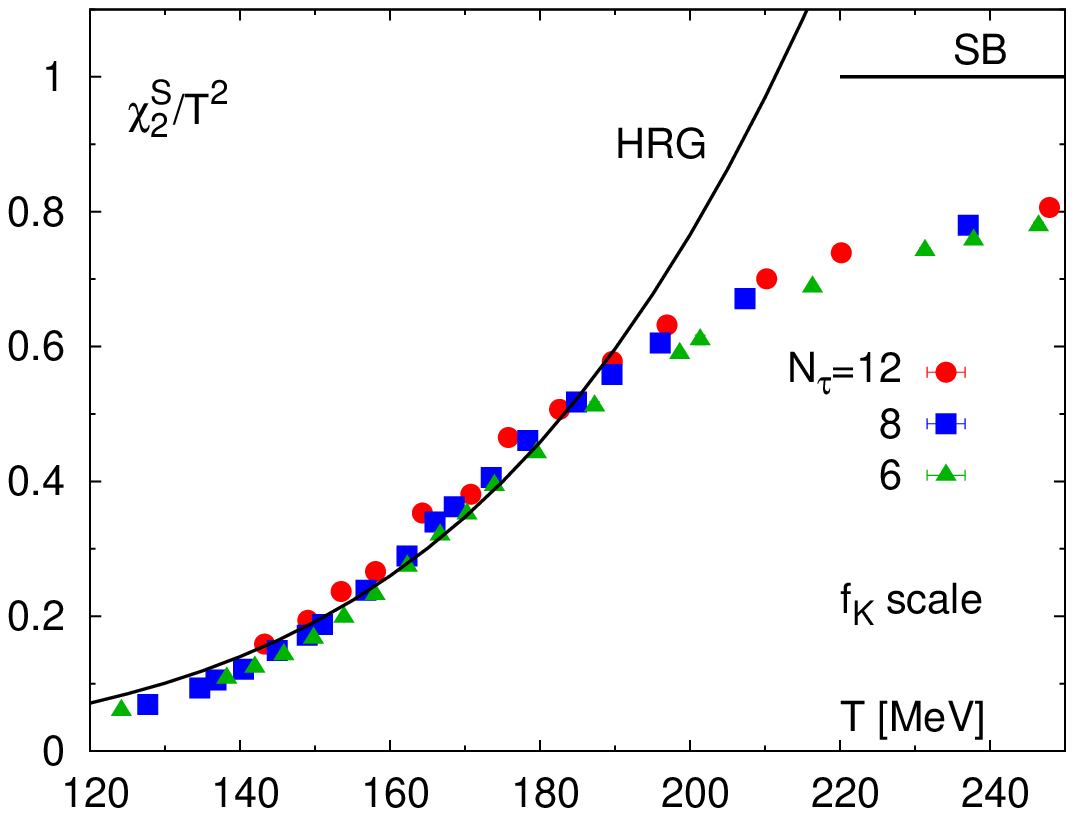,width=7.5cm}\hspace*{-0.6cm}
\epsfig{file=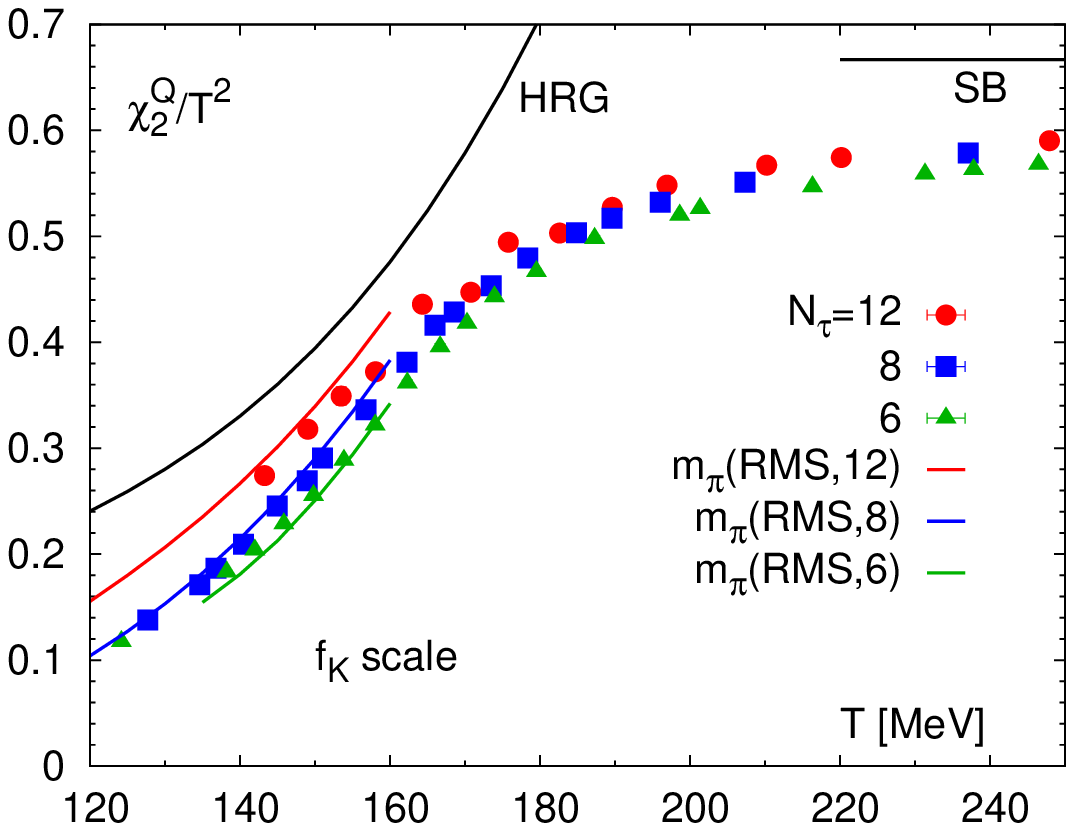,width=7.5cm}
\end{center}
\caption{
Net strangeness (left) and net electric-charge (right) fluctuations versus 
temperature obtained from lattice calculations at three different values
of the lattice cut-off $aT=1/N_\tau$. 
Calculations of $f_K$ have been used to fix the temperature scale.
Also shown are the results obtained from
the HRG model and the infinite temperature ideal gas limit (solid lines)
as well as three short, solid lines for a HRG model in which the pion mass 
has been
replaced by the RMS pion mass relevant for our calculations on
the $N_\tau=6$, 8 and 12 lattices, respectively.
}
\label{fig:chi2QS}
\end{figure}

In Fig.~\ref{fig:chi2QS}, we show results for fluctuations of net
strangeness (left) and net electric charge (right) using $f_K$ to set
the scale for the temperature. The figure also shows the corresponding
HRG results.  Both, the strangeness and 
the baryon number (see Fig.~\ref{fig:chi2B}) fluctuations 
agree with HRG results for temperatures below $T\simeq 160$~MeV and
show small cut-off effects.  The electric charge fluctuations deviate
significantly from HRG results at all temperatures and show large
cut-off dependence. These large cut-off effects in $\hchi_2^Q$ can
mostly be explained as due to discretization errors in the lattice 
hadron spectrum.  The dominant contribution to
$\hchi_2^Q$ at low temperatures comes from pions, while $\hchi_2^S$
receives the leading contribution from kaons and $\hchi_2^B$ from
nucleons. As discussed in Section~\ref{sec:lattice}, the pion spectrum is
strongly affected by taste symmetry violations in the staggered
formulation, and it has been shown in earlier work that the agreement
in $\hchi_2^Q$ between lattice QCD results and HRG model calculations
can be improved using a distorted spectrum in the HRG
analysis~\cite{Tawfik,Petreczky}.  

To confirm this observation, we show in Fig.~\ref{fig:chi2QS} (right
panel) the HRG results obtained after replacing the physical pion mass
by the RMS pion mass relevant to our calculations on the $N_\tau=6$, 8
and 12 lattices, respectively.  These cut-off effects leading to the
spectral distortion vary not only with $N_\tau$ but also with
temperatures at fixed $N_\tau$. We therefore parametrize the cut-off
dependence of $M_\pi^{RMS}$, at each $N_\tau$, using a cubic polynomial
fit to the data given in~\cite{hotQCDTc}. This allowed us to estimate
the modified HRG result as a function of temperature, as shown in
Fig.~\ref{fig:chi2QS} by short solid lines (colored lines).  The data
confirm that below $T\simeq 155$~MeV, the numerical results for
$\hchi_2^Q$ are well described by this minimally modified HRG model
and the major part of the difference is indeed due to the taste
symmetry breaking effects in the pion sector.  Since all other heavier
states contributing to the HRG model are assumed to take on
their physical values, the fluctuations of strangeness and baryon
number are not influenced by this modification of the HRG model.  We
discuss these features, together with continuum extrapolations, for
the data shown in Fig.~\ref{fig:chi2B} and \ref{fig:chi2QS} in more
detail in the following subsection.

It is worthwhile to clarify our discussion of the comparison of
lattice QCD data with the HRG model. When we say that the HRG is a
good approximation to QCD, we refer to the value of the
susceptibility, as is traditional.  At the temperature where this
agreement fails, we observe that not only the value but the slope also
deviates significantly. Our data also indicate that the curvature
starts to deviate 20--30 MeV earlier depending on the observable.
These derivatives of the susceptibility are related to higher moments,
which are increasingly less well captured in the HRG analysis and have
not been calculated in our lattice simulations.

\subsection{Continuum extrapolation and approach to the hadron resonance gas estimates}

In this section, we analyze cut-off effects at fixed values of the
temperature for three values of the lattice spacing, $aT=1/6,\ 1/8$
and $1/12$. We perform this analysis for temperature scales defined in
terms of both $r_1$ and $f_K$ in order to quantify systematic effects
at a given lattice spacing and to demonstrate consistency between the
estimates in the continuum limit. Our simulations on the lattices with the 
three different $N_\tau$ values have not been done at the same values of
the temperature. As a result, in order to perform
continuum extrapolations at fixed temperature, we have used cubic spline
interpolations of our data throughout this paper. We 
propagate errors on the spline parameters. 
\fk{
In addition to estimating statistical errors based on our entire
data sample, we also use the difference of spline interpolations 
performed on two independent sub-sets constructed by choosing 
every second T-point (even and odd points), as 
an estimate for the systematic errors. The
final error on our data is obtained by adding this error estimate
and the purely statistical error obtained from the full data set
in quadrature.
}

\begin{figure}[t]
\begin{center}
\epsfig{file=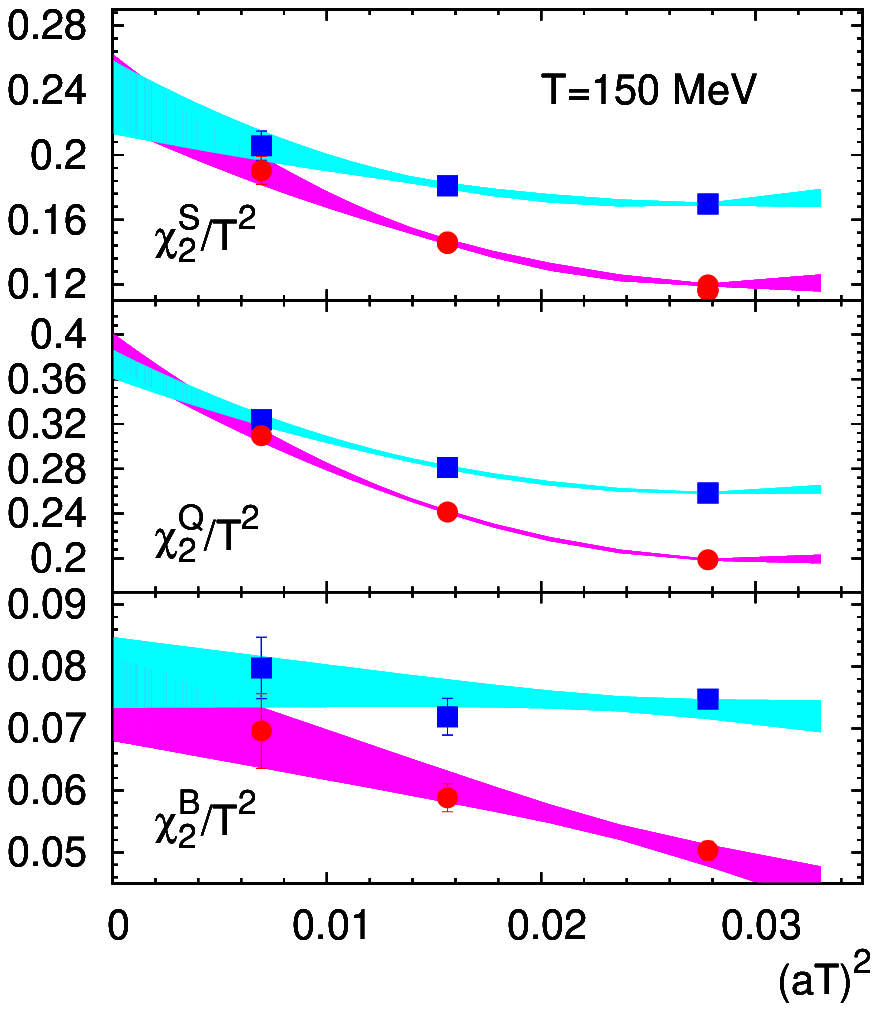,width=7.2cm}
\epsfig{file=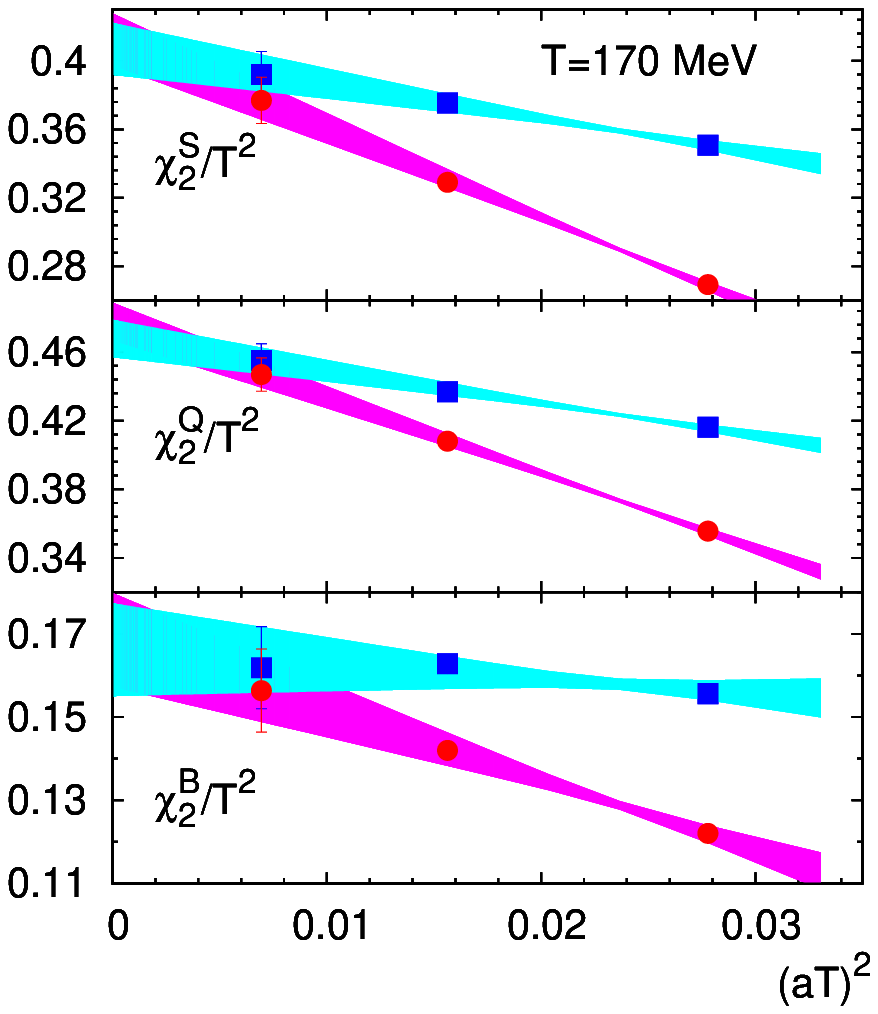,width=7.2cm}

\epsfig{file=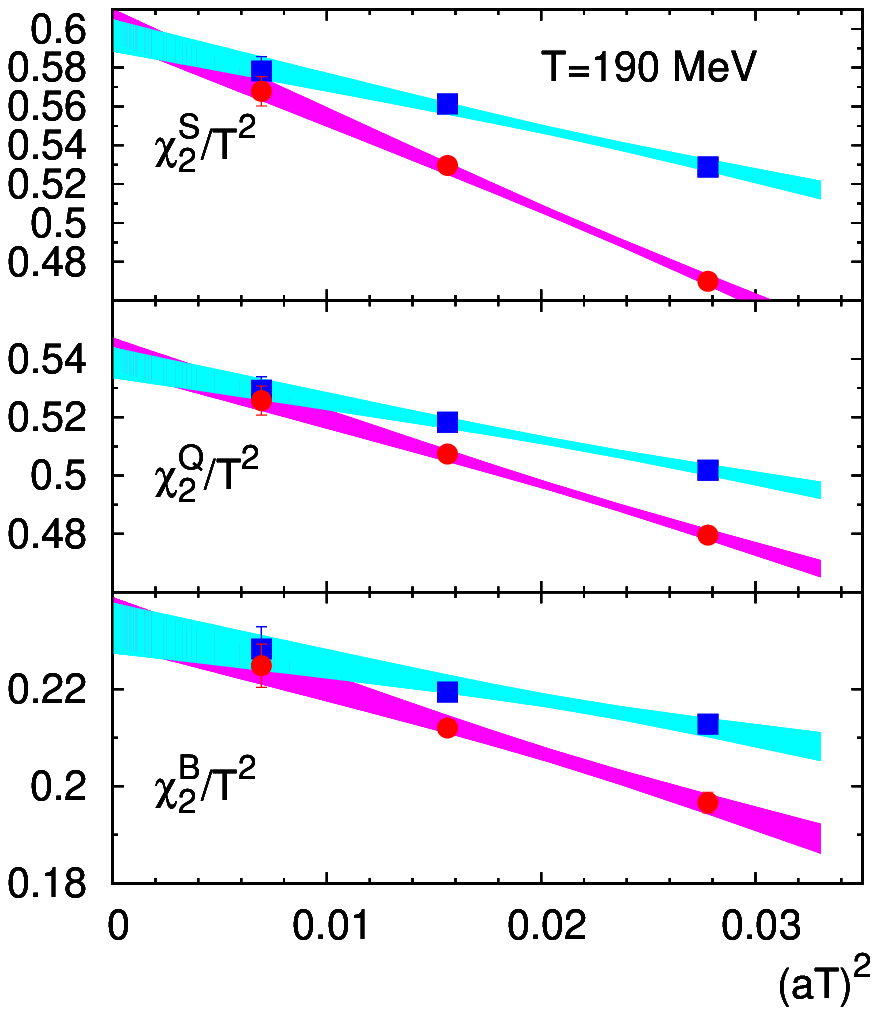,width=7.2cm}
\epsfig{file=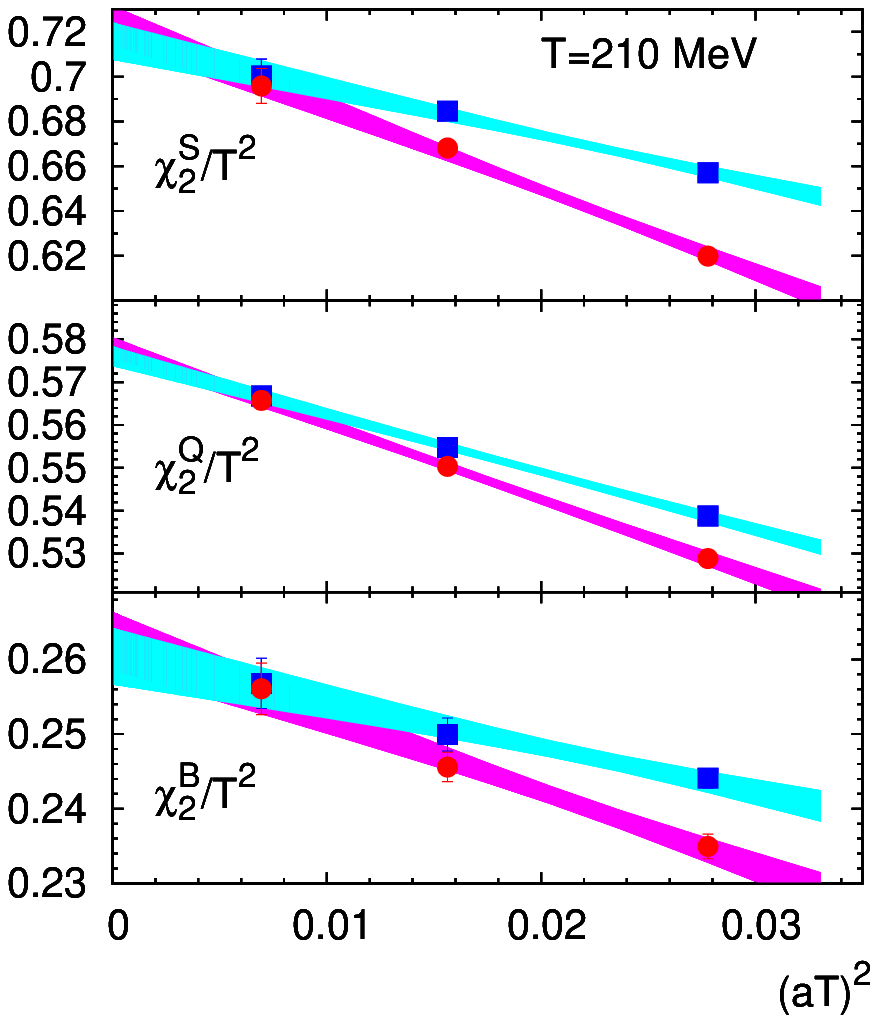,width=7.2cm}
\end{center}
\caption{Continuum extrapolations of net strangeness ($\chi_2^S/T^2$),
net electric charge ($\chi_2^Q/T^2$) and net baryon number ($\chi_2^B/T^2$)
at four values of the temperature using fits linear 
\fk{and, for $\chi_2^S/T^2$ and $\chi_2^Q/T^2$, at the lowest temperature 
quadratic in $(aT)^2=1/N_\tau^2$. 
}
Data at fixed values of the temperature are obtained from cubic spline
interpolations.
The temperature scale has been determined using calculations of 
$r_1$ (circles) and $f_K$ (boxes), respectively.
}
\label{fig:chi2_fK_r1}
\end{figure}

We first discuss the continuum extrapolation for the three
susceptibilities shown in Figs.~\ref{fig:chi2B} and \ref{fig:chi2QS}
at high temperature, $T\ge 170 $~MeV. In this regime, cut-off effects
are generally small, which, to some extent, is due to the fact that
our numerical calculations have been performed with an ${\cal O}(a^2)$
improved action with small cut-off dependence of thermodynamic
observables in the infinite temperature ideal gas limit \cite{Sturm}.
In Fig.~\ref{fig:chi2_fK_r1}, we show continuum extrapolations for all
three susceptibilities, $\chi^{S,Q,B}_2$, at three representative
values of the temperature, $T=170$, $190$ and $210$~MeV. In each case,
and for both temperature scales, $r_1$ and $f_K$, the fits show that the
cut-off effects are consistent with ${\cal O}(g^2(aT)^2)$ corrections
and, over the limited range of $T$, all three susceptibilities can be
extrapolated to the continuum with an Ansatz that includes corrections
linear in $1/N_\tau^2 = (aT)^2$. The continuum extrapolated results
obtained with the two temperature scales agree within errors, and the
results obtained on the $N_\tau=12$ lattices are a good approximation
to these. 
\fk{
This also is true for lower temperatures. However, in this case 
extrapolations linear in $1/N_\tau^2$ are no longer sufficient
for the electric charge and strangeness fluctuations. Systematic
effects at ${\cal O}((aT)^4)$ start to become important. 
This is evident from the data sets at $T=150$~MeV, which are 
shown in Fig.~\ref{fig:chi2_fK_r1} as well. We note that also at this 
temperature, which is the lowest temperature for which we perform 
continuum extrapolations, extrapolations based on the $r_1$ and $f_K$
temperature scales are in good agreement.
}
Having demonstrated consistency of the continuum estimate 
obtained using $r_1$ and $f_K$, we, as stated previously, use the scale from $f_K$ in the
rest of the paper because the slope in the fits is smaller.

\begin{figure}[t]
\begin{center}
\hspace*{-0.6cm}\epsfig{file=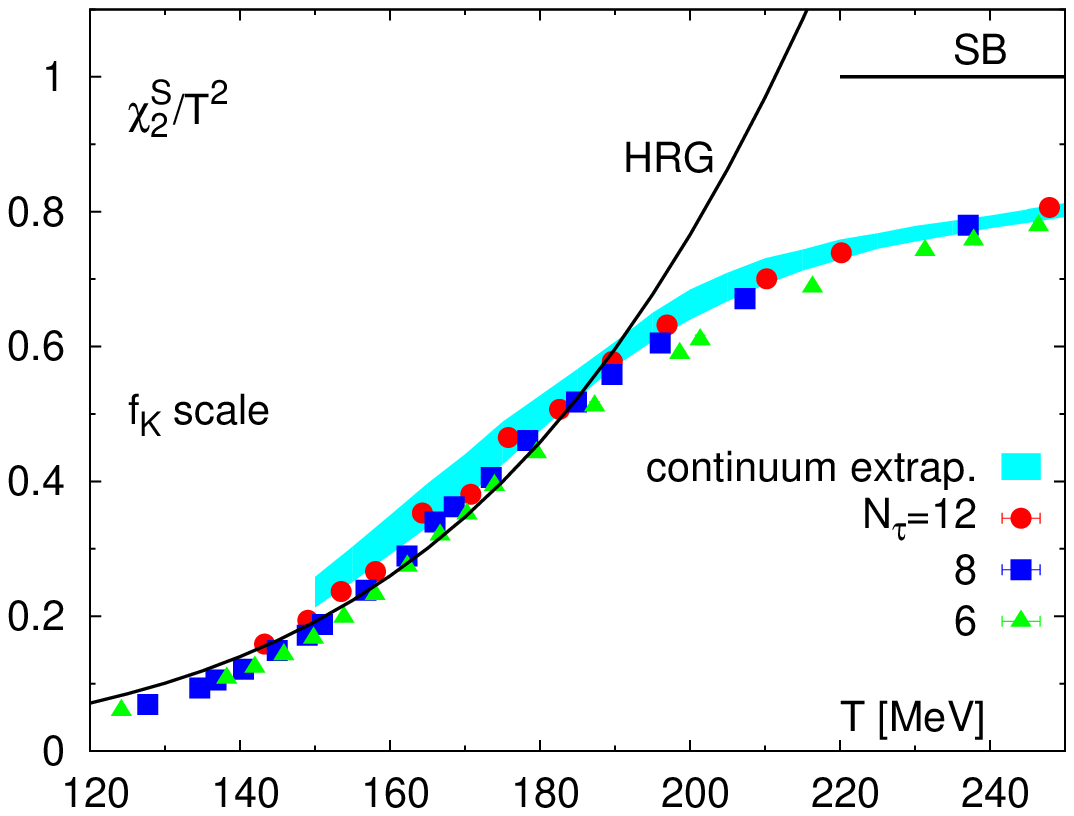,width=7.5cm}\hspace*{-0.6cm}
\epsfig{file=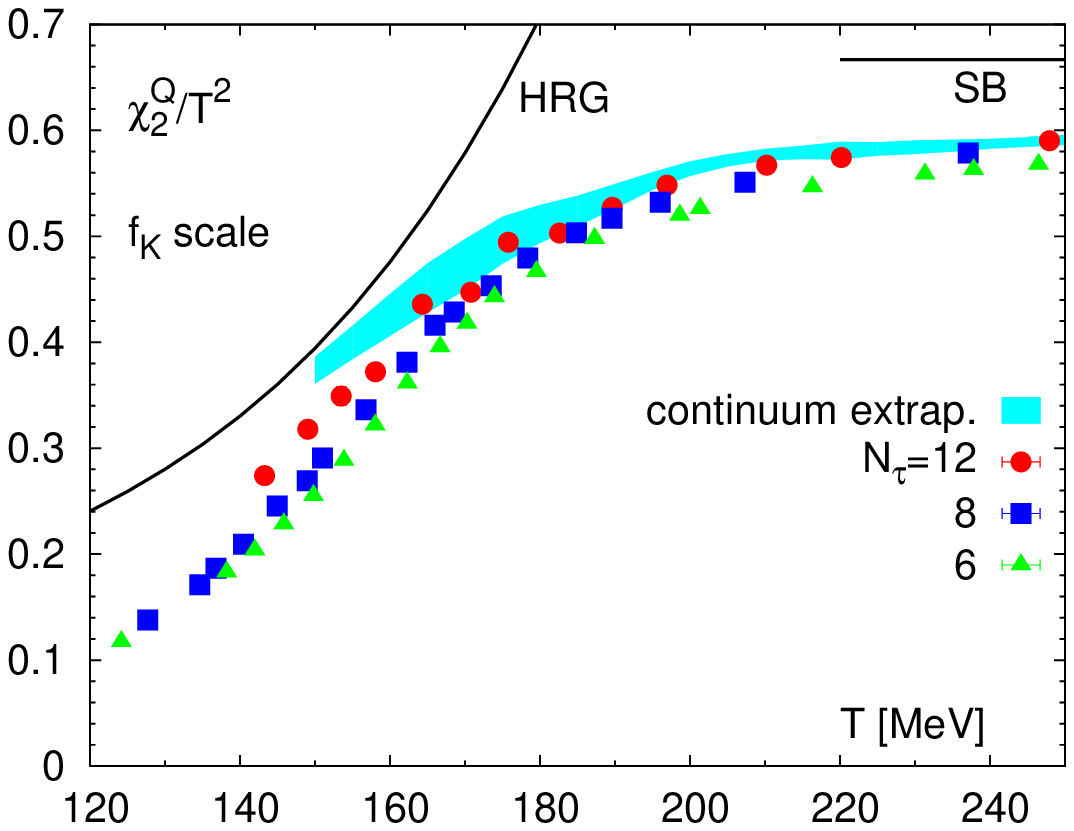,width=7.5cm}
\end{center}
\caption{Fluctuations of net strangeness (left) and electric charge
  (right) in units of $T^2$.  Calculations of $f_K$ have been used to
  fix the temperature scale.  
\fk{Also shown are continuum extrapolated results
  taking into account cut-off effects up to quadratic order in
  $1/N_\tau^2$.
}
  The HRG model result and the SB limit is
  given by the solid lines. }
\label{fig:chi2Q_fits}
\end{figure}

\begin{figure}[t]
\begin{center}
\epsfig{file=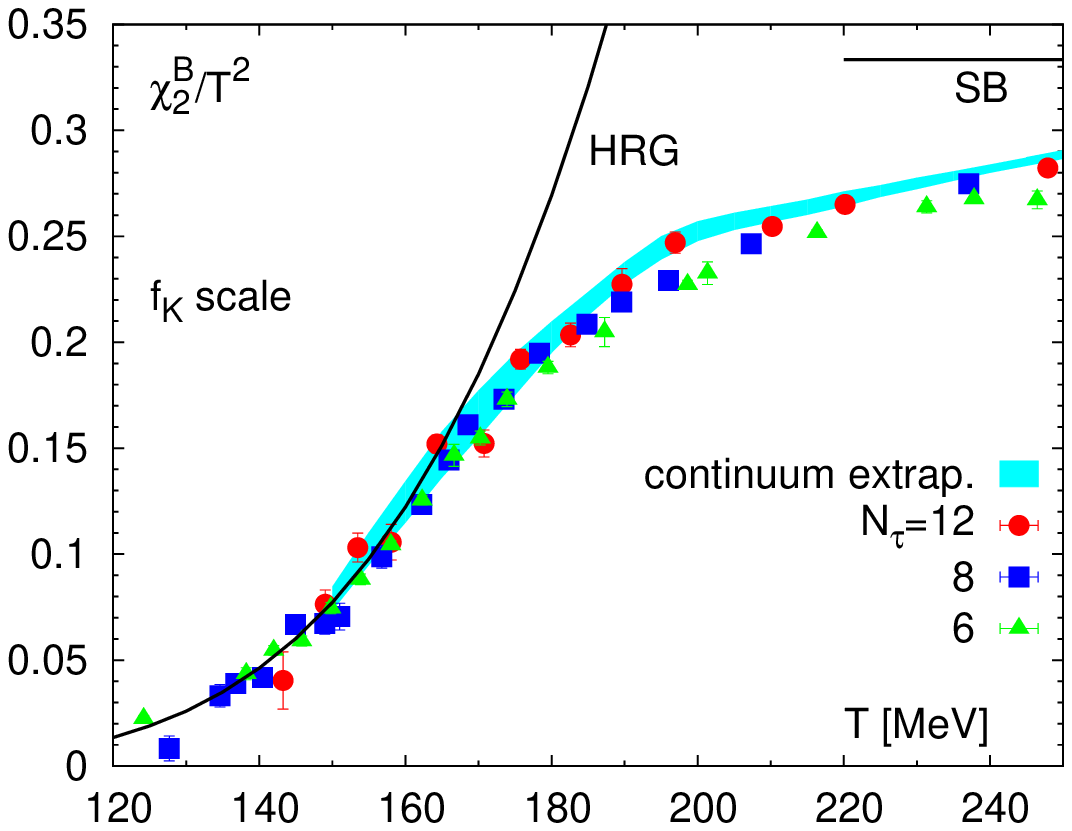,width=7.5cm}
\epsfig{file=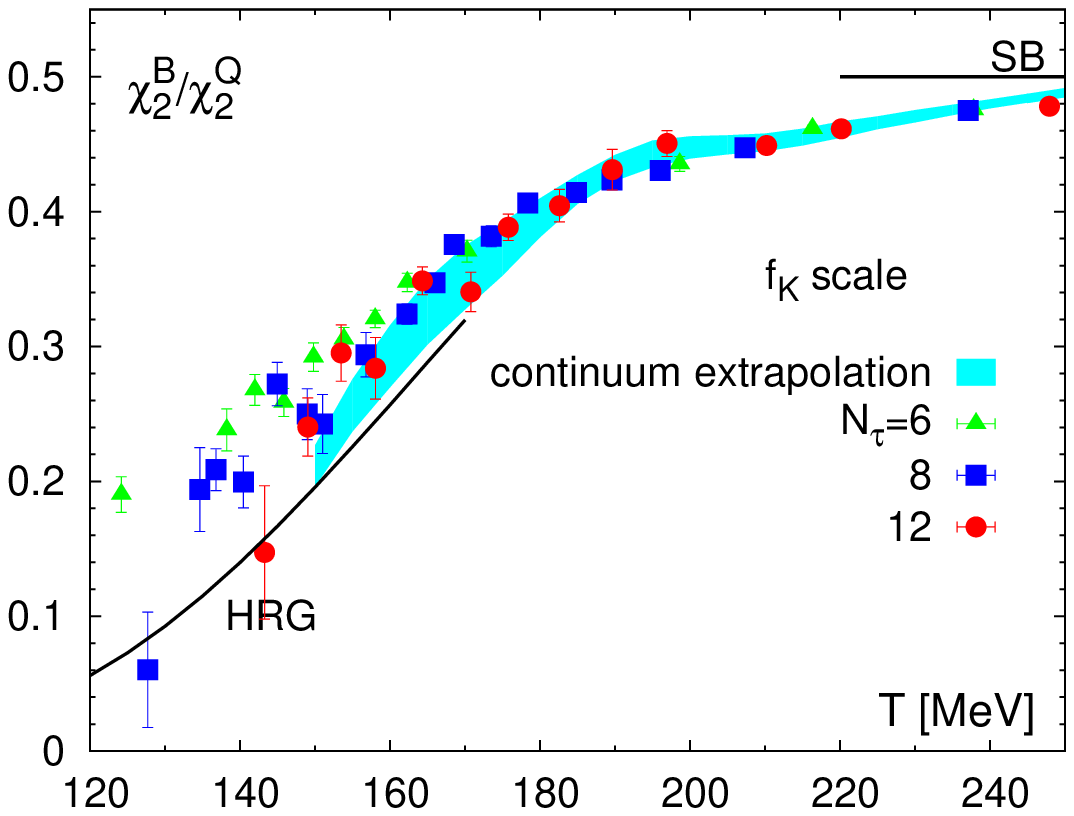,width=7.5cm}
\end{center}
\caption{Net baryon number fluctuations in units of $T^2$ (left)
and the ratio of net baryon number and net electric charge
fluctuations (right).
Calculations of $f_K$ have been used to fix the temperature scale.
Also shown are results from a continuum extrapolation
taking into account ${\cal O}(a^2)$ corrections. 
\fk{For $\chi_2^B/\chi_2^Q$ we also show the ratio
of continuum extrapolations constructed for $\chi_2^B/T^2$
and $\chi_2^Q/T^2$ separately.
}
The HRG model result and the SB limit is
  given by the solid lines.
}
\label{fig:chi2B_fits}
\end{figure}

The data for the net charge fluctuations in the temperature interval
120--250~MeV, results of the 
\fk{linear extrapolation for $\chi_2^B/T^2$,
and quadratic extrapolations for $\chi_2^S/T^2$ and $\chi_2^Q/T^2$ 
}
are shown in Figs.~\ref{fig:chi2Q_fits} and \ref{fig:chi2B_fits}. 
In Fig.~\ref{fig:chi2B_fits}(right) we also show the ratio of net
baryon number and electric charge fluctuations. 
\fk{The continuum
extrapolation shown for this quantity has been obtained from
the corresponding extrapolations for $\chi_2^B/T^2$ and $\chi_2^Q/T^2$.
}

\begin{figure}[t]
\begin{center}
\epsfig{file=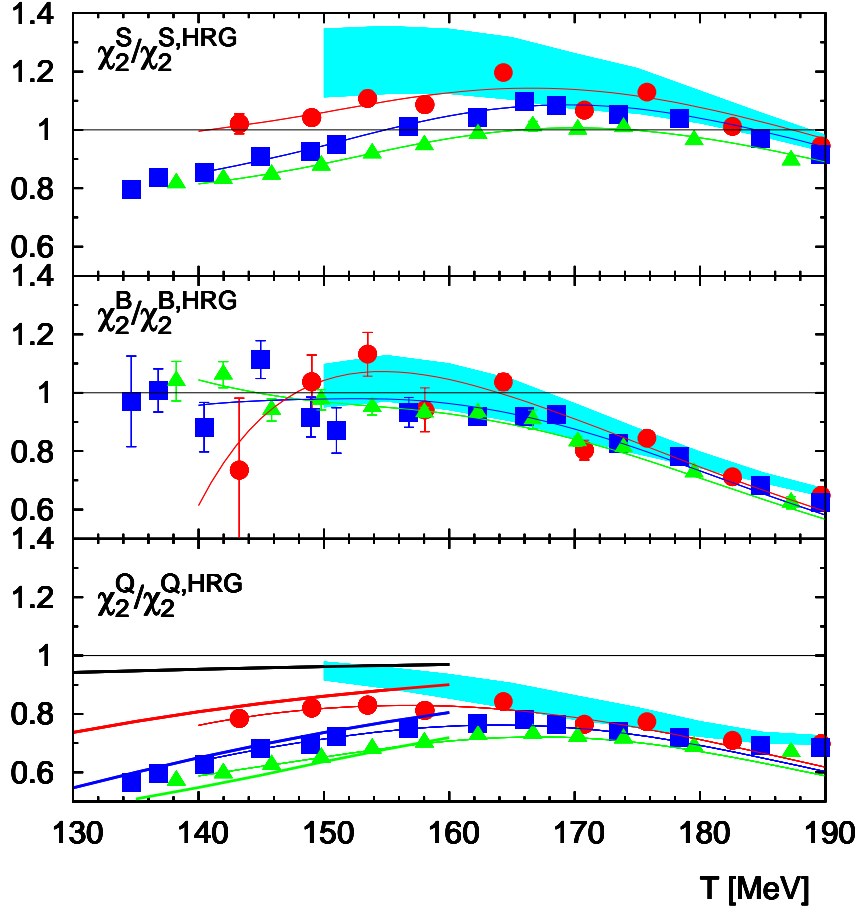,width=15cm}
\caption{Deviations of $\chi_2^{S,B,Q}$ from HRG model results.  Data
  on $N_\tau=6$, $8$ and $12$ lattices are shown using triangles,
  boxes and circles, respectively. Thin lines are the result of a
  cubic spine interpolation of the data at fixed $N_\tau$. 
\fk{
The broad bands are the result of the continuum extrapolations.
}
In the bottom panel, the thick black line just below
  unity is the ratio HRG($M_\pi=160$~MeV)/HRG($M_\pi^{\rm
    physical}$). The other thick lines show the ratio HRG($M_\pi^{\rm
    RMS}$)/HRG($M_\pi^{\rm physical}$) for each of the three $N_\tau$
  lattices.  }
\label{fig:multi}
\end{center}
\end{figure}

Continuum extrapolations in the crossover and the low temperature
regions require additional considerations because the three different
conserved charge susceptibilities show different sensitivities to
cut-off effects.  In order to quantify differences from the HRG model
results in this temperature regime, and in order to clarify the 
extent to which the HRG model provides a good description of QCD results, we
analyze the ratios $\chi_2^X/\chi_2^{X,HRG}$, $X=B,\ Q,$ and $S$, in
Fig.~\ref{fig:multi}.  We find that, while baryon number fluctuations start to agree
with HRG model results for $T\lsim 165$~MeV, the net strangeness
fluctuations become larger than the HRG values for temperatures below
$T\simeq 190$~MeV and then approach the HRG values from above at $T \lsim
150$~MeV. At $T\sim 150$~MeV, the differences are still (10-20)\%.

The electric charge fluctuations show much larger deviations from
the HRG model as is evident from Fig.~\ref{fig:multi}. In particular,
below $T\simeq 170$~MeV, the cut-off dependence in
$\chi_2^Q/\chi_2^{Q,HRG}$ is large and extrapolations including just
leading order $a^2$-corrections fail. As discussed in
Section~\ref{ssec:fluctuations}, this, to a large extent, is due to
the severe cut-off dependence of the pion spectrum, $i.e.$, the
anomalously large RMS pion mass suppresses fluctuations in the
electric charge and has a much smaller effect on the baryon and strangeness
charges. In short, a continuum extrapolation without including the
effects of taste symmetry breaking is insufficient.

The distorted HRG model, which modifies the log of the partition
function by replacing $M_\pi$ by $M_\pi^{\rm RMS}$ in the pion
contribution, $\exp(-M_\pi/T)$, however, does describe the data
well. In general, the HRG model defined in Eq.~\ref{ZMB} suggests that, in this
temperature regime, cut-off effects in any quantity $f$ may be
accounted for by an exponential Ansatz of the form
\begin{equation}
f(N_\tau,T) = a(T) +b(T)\ {\rm e}^{-c(T)/N_\tau^2} \; ,
\label{HRG_fit}
\end{equation}
which, at high temperatures where cut-off effects become small, 
reduces to the linear fit in $1/N_\tau^2$, {\it i.e.,}
$f(N_\tau,T) \simeq \tilde{a}(T)+ \tilde{b}(T)/N_\tau^2$
\fk{and also incorporates the next to leading order
quadratic corrections
$f(N_\tau,T) \simeq \tilde{a}(T)+ \tilde{b}(T)/N_\tau^2+
\tilde{c}(T)/N_\tau^4$.
}
We, therefore, analyze our data for $\hchi_2^{Q,S}$ in the transition
region, $150 \le T\le 190~{\rm MeV}$, from the low to high temperature
phase of (2+1)-flavor QCD using fits linear and quadratic in
$1/N_\tau^2$ as well as the exponential Ansatz given in
Eq.~(\ref{HRG_fit}).  
\fk{With our current statistical accuracy we are, however, not
sensitive to cut-off effects beyond ${\cal O}((aT)^4)$. In fact,
all our fits performed with the exponential ansatz are
consistent within errors with fits based on the quadratic 
ansatz. We, therefore, do not discuss the exponential fits any 
further in this paper.
In the case of $\hchi_2^{B}$ we find 
that linear and quadratic 
fits agree within errors  
and lead to a $\chi^2/dof$
less than unity in the entire range of temperatures $T\ge 150$~MeV. 
We therefore use
the linear fits to perform continuum extrapolations for $\hchi_B$.
For strangeness and electric charge susceptibilities we use 
quadratic extrapolations in the entire temperature range, although,
as discussed above, we do not observe systematic differences between 
linear and quadratic extrapolations for $T\gsim 170$~MeV. Using
the latter for our continuum extrapolations, however, leads to more
conservative error estimates.
} 

\fk{A summary of our continuum extrapolations for
$\chi_2^X/\chi_2^{X,HRG}$ 
and the data in the low temperature region
is also shown in Fig.~\ref{fig:multi}.  
Due to taste symmetry breaking, the data show significant dependence on $N_\tau$ 
for $T\lsim 170$~MeV.}
To understand this cut-off effect we compare, in the bottom panel in Fig.~\ref{fig:multi}, results for 
$\chi_2^Q/\chi_2^{Q,HRG}$ with 
a modified HRG model in which the physical pion mass is replaced by (i) the
$N_\tau$ dependent RMS mass, HRG($M_\pi^{\rm RMS}$)/HRG($M_\pi^{\rm
  physical}$), and (ii) a pion mass of $160$~MeV,
HRG($M_\pi=160$~MeV)/HRG($M_\pi^{\rm physical}$), corresponding to the
light quark mass actually used in our calculations.  Since the line
showing HRG($M_\pi=160$~MeV)/HRG($M_\pi^{\rm physical}$) is much
closer to unity compared to the other three, we confirm that the
errors due to simulating at this slightly heavier pion mass are 
significantly smaller than the cut-off effects leading to a much
heavier RMS mass. 


In the interesting temperature range relevant to the discussion of
freeze-out conditions in heavy ion collisions, $160~{\rm MeV} \lsim T
\lsim 170~{\rm MeV}$, we find that the continuum extrapolated electric
charge fluctuations are (10-20)\% smaller than even the modified HRG
model calculation with $M_\pi=160$~MeV. For temperatures $T\lsim
150$~MeV, the $\chi_2^{Q}$ data start to agree with the modified HRG
results with $M_\pi=M_\pi^{\rm RMS}$ and continuum extrapolations using the
quadratic Ansatz start to agree with the HRG result.

Strangeness fluctuations on the other hand, both for the $N_\tau=12$
data and the continuum extrapolated values, stay systematically above
the hadron resonance gas result in the temperature range 150--190~MeV.
We, therefore, expect this feature to survive the continuum
extrapolation.  Below $T\lsim 150$~MeV, fluctuations in the
strangeness charge show an $N_\tau$ dependence, which is most
likely again due to taste symmetry violations.  More data are required
to study this issue further. 

Net baryon number fluctuations are consistent with HRG model results
for temperatures below $T\simeq 160$~MeV, although statistical errors
on our $N_\tau=12$ data set make the quantification of possible
deviations from the HRG result in this temperature range difficult.
For larger values of the temperature, the estimates and errors in them
grow progressively smaller.  Our linear extrapolations suggest that in
the temperature interval $160~{\rm MeV} \lsim T \lsim 170~{\rm MeV}$, 
deviations from the HRG model calculations are at most 10\%.
A confirmation of this in our data, through the inclusion of quadratic
corrections, however, requires better statistics.   
Data obtained with the stout action \cite{Fodor_11} also
suggest that the $\chi_2^B$ stays close to the HRG model result in this
temperature range.

In order to reflect the influence of systematic effects on our 
continuum extrapolation we varied fit ranges and distribution of knots
in the smooth spline interpolation. Moreover, in order to account for
possible underestimates of errors on the individual data points, 
\fk{we divided our data samples into two independent sub-sets
consisting of even and odd T-values by rank-order. 
We used the 
}
differences in these fits and the fit to the full data sample as an additional error on 
our spline interpolations.
 

To summarize, our continuum extrapolated values of $\chi_2^{B,Q,S}/T^2$ 
are given  in Table~\ref{tab:all_fluctuations}. 
\fk{Extrapolations of
$\chi_2^{Q,S}/T^2$ used the Ansatz, Eq.~(\ref{HRG_fit})
truncated at
$O(1/N_\tau^4)$, i.e., we used the quadratic extrapolations, whereas
for $\chi_2^{B}/T^2$ the exponential was truncated at
$O(1/N_\tau^2)$, i.e., we used the linear extrapolations. 
For $\chi_2^{Q,S}/T^2$ we also find that the quadratic extrapolated 
values agree with the exponential Ansatz. Our extrapolated results are 
in good agreement with the recently published analysis using the stout 
action \cite{Fodor_11}.}

\begin{table}[t]
 \begin{tabular}{|c|ccc|ccc|}
 \hline
 $T$ [MeV]& $\chi_2^B/T^2$  & $\chi_2^Q/T^2$ & $\chi_2^S/T^2$ & $\chi_{11}^{BS}/T^2 $ & $\chi_{11}^{BQ}/T^2$ & $\chi_{11}^{QS}/T^2$ \\
 \hline
150 &   0.0790( 57)  &   0.3736(126)  &   0.2358(225)  &  -0.0415( 60)  &   0.0187( 29)  &   0.0929( 61) \\
155 &   0.1026( 77)  &   0.3998(158)  &   0.2770(260)  &  -0.0587( 69)  &   0.0221( 24)  &   0.1031( 68) \\
160 &   0.1247( 98)  &   0.4258(198)  &   0.3208(289)  &  -0.0767( 75)  &   0.0242( 23)  &   0.1173( 70) \\
165 &   0.1470(109)  &   0.4513(232)  &   0.3645(322)  &  -0.0949( 81)  &   0.0259( 21)  &   0.1332( 78) \\
170 &   0.1662(111)  &   0.4734(240)  &   0.4067(334)  &  -0.1115( 84)  &   0.0269( 21)  &   0.1491( 84) \\
175 &   0.1849( 99)  &   0.4964(222)  &   0.4567(314)  &  -0.1304( 79)  &   0.0270( 19)  &   0.1655( 83) \\
180 &   0.2024( 74)  &   0.5118(179)  &   0.5013(259)  &  -0.1497( 66)  &   0.0263( 17)  &   0.1782( 77) \\
185 &   0.2182( 56)  &   0.5239(144)  &   0.5449(215)  &  -0.1682( 57)  &   0.0249( 19)  &   0.1892( 76) \\
190 &   0.2326( 52)  &   0.5379(115)  &   0.5888(181)  &  -0.1860( 56)  &   0.0233( 22)  &   0.2005( 72) \\
195 &   0.2445( 55)  &   0.5522( 84)  &   0.6289(210)  &  -0.2005( 60)  &   0.0220( 21)  &   0.2122( 68) \\
200 &   0.2524( 46)  &   0.5638( 68)  &   0.6623(222)  &  -0.2116( 52)  &   0.0205( 21)  &   0.2235( 70) \\
205 &   0.2570( 41)  &   0.5717( 57)  &   0.6884(210)  &  -0.2200( 38)  &   0.0186( 18)  &   0.2335( 67) \\
210 &   0.2604( 38)  &   0.5770( 55)  &   0.7113(194)  &  -0.2267( 31)  &   0.0168( 15)  &   0.2424( 62) \\
215 &   0.2637( 36)  &   0.5792( 63)  &   0.7282(155)  &  -0.2335( 32)  &   0.0152( 14)  &   0.2486( 51) \\
220 &   0.2676( 35)  &   0.5809( 84)  &   0.7438(148)  &  -0.2397( 35)  &   0.0140( 12)  &   0.2537( 57) \\
225 &   0.2713( 28)  &   0.5824( 63)  &   0.7565(113)  &  -0.2456( 34)  &   0.0128( 12)  &   0.2576( 56) \\
230 &   0.2749( 27)  &   0.5841( 64)  &   0.7672(116)  &  -0.2511( 33)  &   0.0119( 12)  &   0.2606( 56) \\
235 &   0.2784( 23)  &   0.5855( 56)  &   0.7760(105)  &  -0.2564( 28)  &   0.0109( 10)  &   0.2628( 48) \\
240 &   0.2819( 20)  &   0.5873( 46)  &   0.7847( 92)  &  -0.2613( 24)  &   0.0101(  8)  &   0.2649( 39) \\
245 &   0.2852( 20)  &   0.5890( 45)  &   0.7934( 98)  &  -0.2664( 23)  &   0.0093(  8)  &   0.2667( 37) \\
250 &   0.2885( 20)  &   0.5907( 45)  &   0.8020( 98)  &  -0.2710( 23)  &   0.0085(  7)  &   0.2688( 34) \\
 \hline
 \end{tabular}

\begin{center}
\end{center}
\caption{
Continuum extrapolated results for the quadratic fluctuations of net 
baryon number, electric charge and strangeness densities
\fk{and the correlations among them}.
Results for 
$\chi_2^Q/T^2$ and $\chi_2^S/T^2$ are 
obtained from 
{quadratic fits and those for $\chi_2^B/T^2$ from linear fits.}
}
\label{tab:all_fluctuations}
\end{table}

Lastly, in Fig.~\ref{fig:chi2B_fits} (right), we show the ratio of the
net baryon number and electric charge fluctuations,
${\chi_2^B}/{\chi_2^Q}$.  It approaches the HRG model result from
above and starts to agree with it for $T\lsim 150$~MeV.  The continuum
extrapolation here is based on the linear and exponential
extrapolations for $\chi_2^B$ and $\chi_2^Q$, respectively.  In the
region of interest to heavy ion phenomenology, this ratio varies
between 
\fk{$0.29(4)$ 
}
(at 160~MeV) and 
\fk{$0.35(4)$ 
}
(170~MeV). Thus,
fluctuations in net electric charge could be 3--4 times larger 
than in the net baryon number in the vicinity of the freeze-out temperature.

\section{Correlations}

\begin{figure}[t]
\begin{center}
\epsfig{file=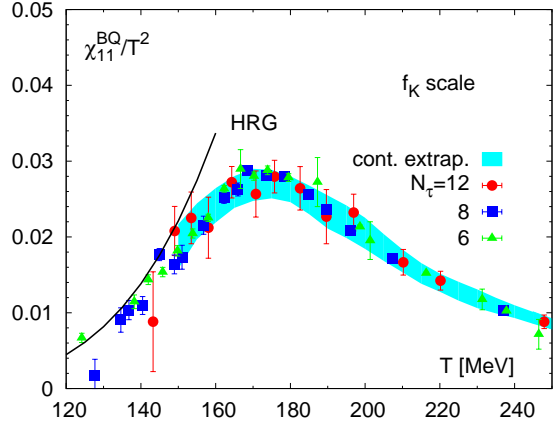,width=8.5cm}
\caption{
Correlations of electric charge with baryon number versus temperature.
The temperature scale has been set using $f_K$. The solid line shows the
result for the HRG model.
\fk{The band shows continuum extrapolations that 
take into account cut-off effects linear in $1/N_\tau^2$.
}
}
\label{fig:chiBQ}
\end{center}
\end{figure}

Probes of the structure of QCD at finite temperature include
correlations among different conserved charges. These correlations
show characteristic changes in the crossover region between the low
and high temperature phases of QCD, which are correlated with changes
in the relevant degrees of freedom.  They also provide insight into
the applicability of HRG model calculations at low temperatures.  The
change in correlations between baryon number and electric charge,
$\hchi_{11}^{BQ}$, is expected to be particularly striking as one goes from the low to
the high temperature phase.  At low temperatures this correlation is
dominated by the contribution of protons plus
anti-protons. Consequently, within the HRG model it rises
exponentially with temperature in this region. In the high temperature
limit of (2+1)-flavor QCD, however, $\hchi_{11}^{BQ}$ vanishes as the
quarks become effectively massless, ($m_i/T\rightarrow 0$), and the
weighted sum of the charges of up, down and strange quarks vanishes.
Results for $\hchi_{11}^{BQ}$ shown in Fig.~\ref{fig:chiBQ} are
consistent with this picture.

\begin{figure}[t]
\begin{center}
\hspace*{-0.6cm}\epsfig{file=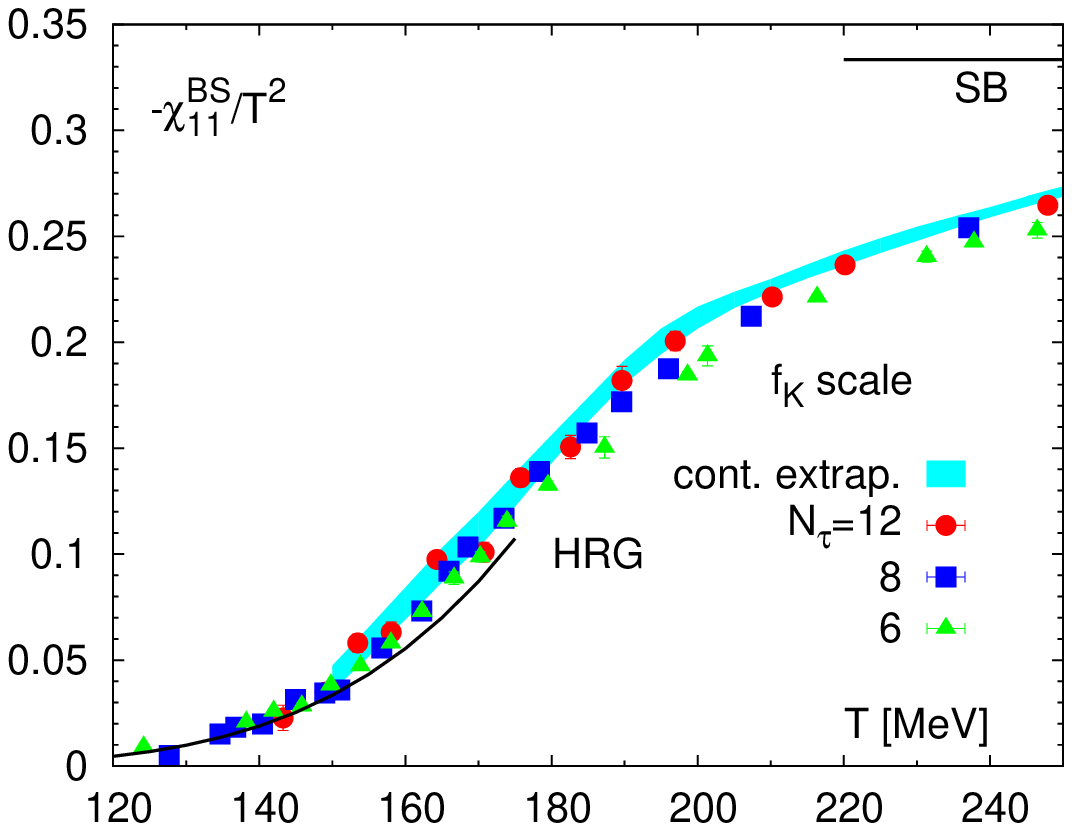,width=7.5cm}\hspace*{-0.6cm}
\epsfig{file=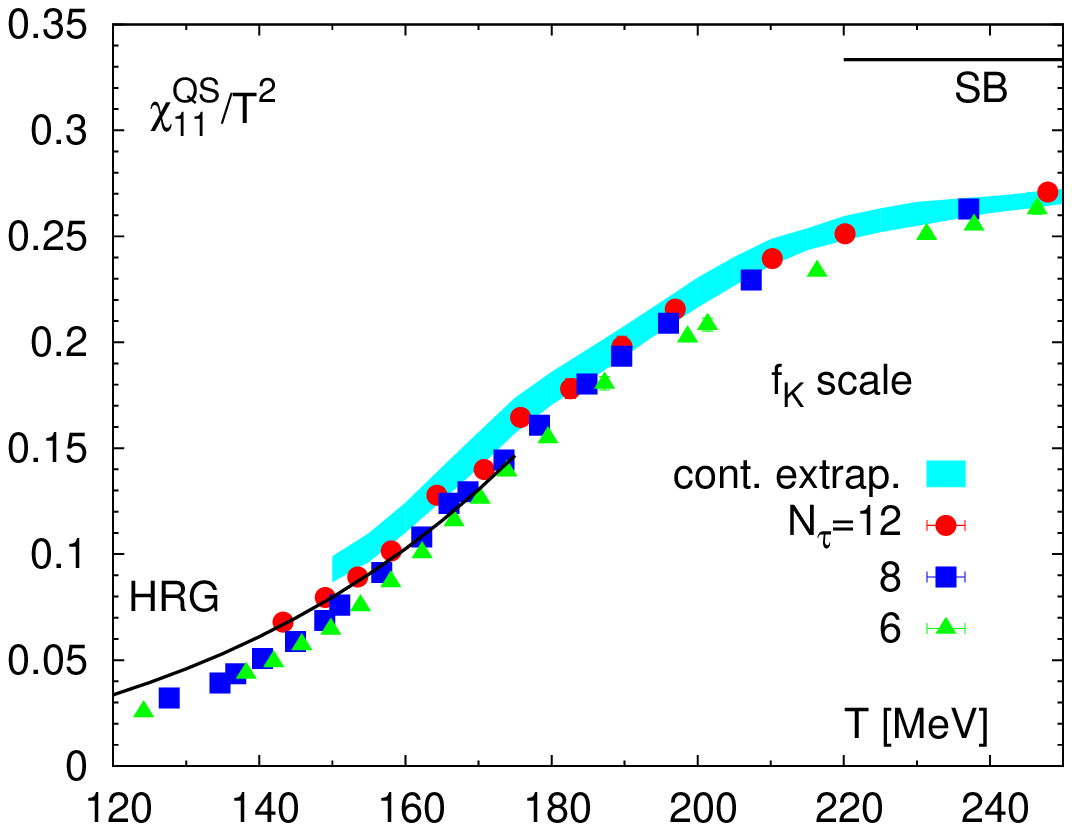,width=7.5cm}
\caption{
Correlations of strangeness with baryon number (left) and 
electric charge (right).
In both cases, $f_K$ has been used to set the scale. The solid lines show the
result for the HRG model. 
\fk{For $\chi_{11}^{BS}/T^2$ the band shows continuum extrapolations that 
take into account cut-off effects linear in $1/N_\tau^2$ while for 
$\chi_{11}^{QS}/T^2$ also quadratic corrections have been accounted for.
}
}
\label{fig:chiBS}
\end{center}
\end{figure}

The correlations of strangeness with the baryon number and the
electric charge, $\hchi_{11}^{BS}$ and $\hchi_{11}^{QS}$, are
sensitive to changes in the strangeness degrees of freedom
\cite{Koch1,Koch2,Majumder}. Results for the temperature dependence of
these correlations are shown in Fig.~\ref{fig:chiBS}. They approach
the Stefan-Boltzmann value, $1/3$, of a massless three flavor quark gas
at high temperatures. As observed in the case of the quadratic
fluctuations, on decreasing the temperature towards the transition
region, these correlations first overshoot the HRG model result and then
approach HRG value from above at about $150$~MeV.  This overshoot is
more pronounced for $-\hchi_{11}^{BS}$ than for $\hchi_{11}^{QS}$.
\fk{
Also shown in Fig.~\ref{fig:chiBS} are continuum extrapolations (bands)
which in the baryon sector, i.e., for B-S and B-Q correlations, are
based on fits linear in $1/N_\tau^2$, whereas in the meson sector, i.e.,
for Q-S correlations, quadratic corrections are also taken into account.
This reflects the larger sensitivity of the latter to taste violations
that also has been observed for the quadratic strangeness and 
electric charge fluctuations.
}

In the isospin symmetric case considered in this study, 
the flavor correlations $\hchi_{11}^{us}$ 
and $\hchi_{11}^{ds}$ are equal. Also, the two correlations
$2 \hchi_{11}^{QS}$ and $\hchi_{11}^{BS}$ 
are related to each other through the quadratic strangeness
fluctuations, {\it i.e.,} $2 \hchi_{11}^{QS} - \hchi_{11}^{BS} =
\hchi_2^S$~\cite{Majumder}. One can then write the following relationships
between the charge correlations and quark-flavor fluctuations:
\begin{eqnarray}
\hchi_{11}^{QS} &=& \frac{1}{3}\left( \hchi_2^s -\hchi_{11}^{us}\right) \;\; , \
\nonumber \\ 
\hchi_{11}^{BS} &=& -\frac{1}{3}\left( \hchi_2^s +2 \hchi_{11}^{us} \right)
\;\; .
\label{QS_BS}
\end{eqnarray}
At high temperatures $\hchi_{11}^{QS} \sim - \hchi_{11}^{BS}$ because
$\hchi_{11}^{us}$ receives perturbative contributions only at ${\cal
  O}(g^6\ln(1/g^2))$ and is therefore small~\cite{Allton,Blaizot}. On
the other hand, corrections to $\hchi_2^s$ from the ideal gas limit are
dominant as they are ${\cal O}(g^2)$.

\begin{figure}[t]
\begin{center}
\hspace*{-0.6cm}\epsfig{file=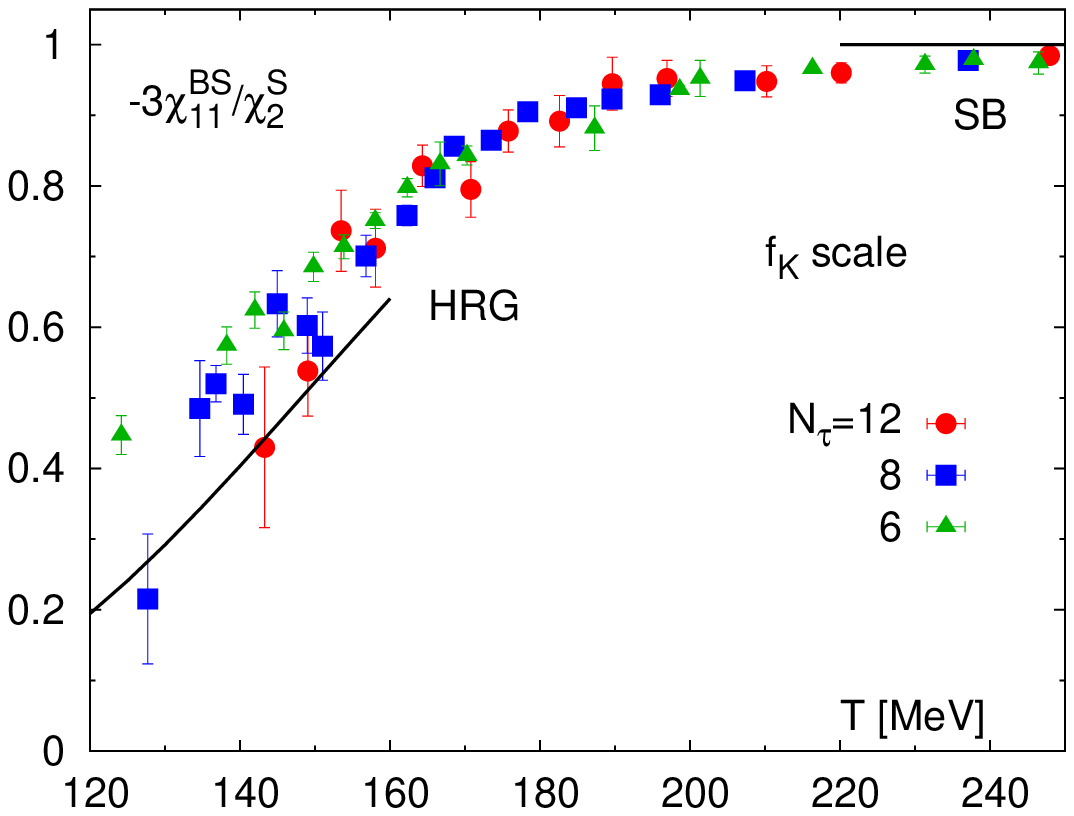,width=7.5cm}\hspace*{-0.6cm}
\epsfig{file=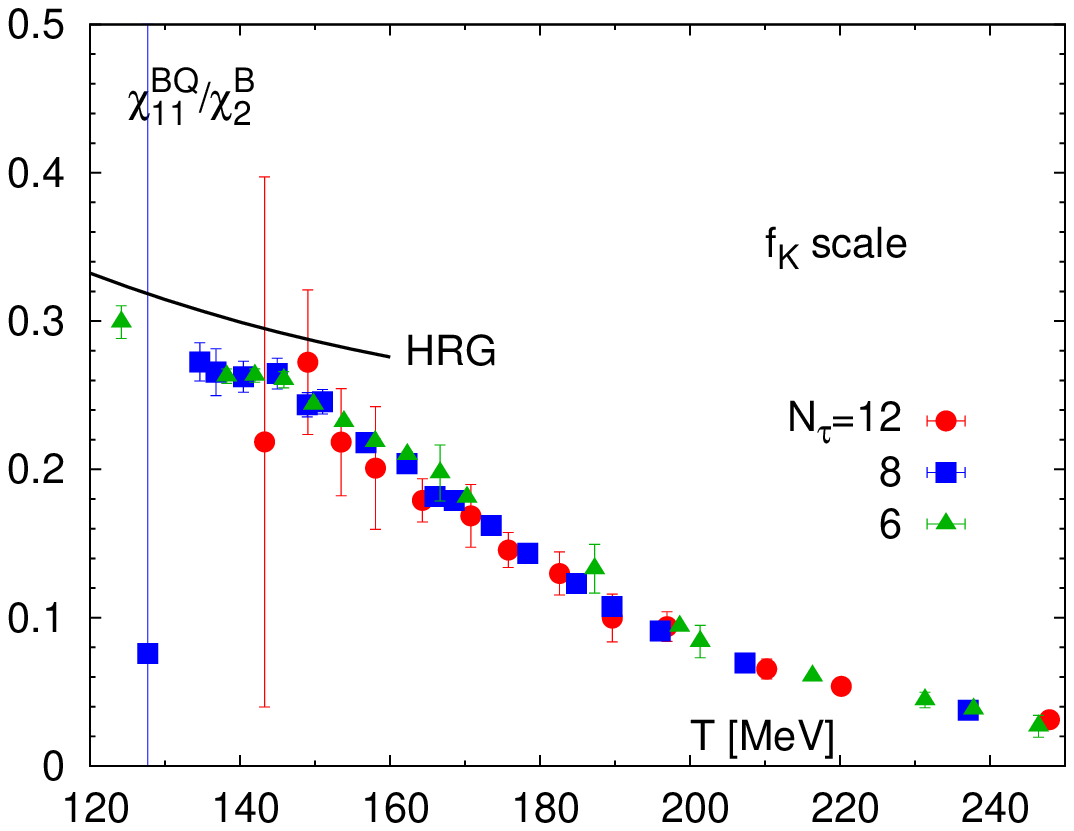,width=7.5cm}
\caption{
Correlations between baryon number and strangeness (left) as well as 
electric charge (right) normalized to strangeness and net baryon number
fluctuations, respectively. The solid line shows the HRG model result.
}
\label{fig:BS_S2}
\end{center}
\end{figure}

All three charge correlations show significant deviations from the
ideal gas limit even at twice the transition temperature 
(see data in Tables~\ref{tab:chi2_6}, \ref{tab:chi2_8}
and \ref{tab:chi2_12}). These deviations are due to large contributions 
of flavor fluctuations, such as to $\hchi_2^s$ discussed above. 
The leading order perturbative correction can be eliminated by
forming suitable ratios that can be used to analyze experimental data
on charge fluctuations~\cite{Koch1,Koch2,Majumder},
\begin{eqnarray}
C_{BS} &=& - 3 \frac{\chi_{11}^{BS}}{\chi_2^{S}}  \;\; , \nonumber \\
C_{QS} &=& 3 \frac{\chi_{11}^{QS}}{\chi_2^{S}} = \frac{1}{2}\left( 3-C_{BS}
\right) \; .
\label{Koch}
\end{eqnarray}
At high temperature the deviations from the ideal gas value of unity
are now due to $\chi_{11}^{us}/\chi^s_2$, for example, $C_{BS} = 1 +
2\chi_{11}^{us}/\chi^s_2$. Data for $C_{BS}$ is shown in
Fig.~\ref{fig:BS_S2}(left) and, in comparison to the
quadratic strangeness fluctuations (Fig.~\ref{fig:chi2QS}) or baryon number strangeness
correlations (Fig.~\ref{fig:chiBS}), the approach to the ideal gas limit is much more
rapid. This shows that the flavor correlation $2 \chi_{11}^{us}/\chi^s_2$ is already 
small for $T\gsim 1.2 T_c$.  It is, however, large in the vicinity
of the transition temperature.

The behavior of the third ratio that one can analyze, $C_{BQ} =
\chi_{11}^{BQ} / \chi_2^{B}$, is somewhat different as leading order
perturbative corrections do not cancel completely due to differences
in the light and strange quark masses. Consequently, the approach to
the ideal gas limit is slower as can be seen in
Fig.~\ref{fig:BS_S2}(right).

\section{Discussion and Conclusion}

We have analyzed quadratic fluctuations and correlations among
conserved charges in (2+1)-flavor QCD. We find that as the temperature
is decreased from the high temperature phase, the net baryon number 
fluctuations start to agree 
with the hadron resonance gas model below $165$~MeV while for electric
charge fluctuations this happens only below $150$~MeV. The fluctuations of
net strangeness overshoot the HRG model values at $T \sim 190$~MeV. In
the temperature range relevant to the discussion of chemical freeze-out
in heavy ion collisions, 160--170~MeV, strangeness fluctuations are
systematically larger than the HRG model result by about 20\%. These
detailed differences between QCD calculations and HRG model results
should become manifest when experimental data on the probability
distributions of net charge fluctuations \cite{STAR} is analyzed. 
In fact, quadratic fluctuations characterize the bulk
structure of these distributions which, in the Gaussian approximation
at $\mu_B=0$, is given by
\begin{equation}
P(N_X) = {\rm e}^{- N_X^2/(2 V_f T^3_f \hchi_2^X)} \;\; ,\;\; X=B,\ S,\ Q \; ,
\label{distribution}
\end{equation}
where $T_f$ denotes the temperature at the time of chemical freeze-out
and $V_f$ is the freeze-out volume. The results presented here suggest
that the largest deviations from HRG model calculations occur in the
probability distributions for electric charge and strangeness
fluctuations.  We give a summary of the fluctuations in conserved charges
for temperatures in the transition region in
Table~\ref{tab:fluctuations}.

In the phenomenologically interesting temperature regime $160~{\rm
  MeV}\lsim T \lsim 170~{\rm MeV}$, continuum extrapolated results for
$\hchi_2^Q$, are smaller than the HRG model results by about
10--20\%. Even though at temperatures below $150$~MeV, estimates for
electric charge fluctuations have large systematic errors due to the
distortion of the light meson spectrum in all staggered formulations,
our analysis shows that these effects can be taken into account when
performing continuum extrapolations. We show that the resulting
continuum estimates in the transition region lie below the HRG
estimates even after corrections accounting for the distorted pion
spectrum have been applied. Thus, our conclusion is that at the
highest RHIC energies and at the LHC, the width of probability
distribution for the net electric charge should be narrower 
than what HRG model calculations would suggest since 
$\hchi_2^Q$ in Eq.~\ref{distribution} is smaller. 

In the case of the 
net baryon number fluctuations, deviations from HRG model results 
start to become statistically significant 
only for $T\gsim 165$~MeV and will therefore be hard to quantify. 
We also find that the ratio of the net baryon number and electric charge fluctuations, presented in 
Fig.~\ref{fig:chi2B_fits}, approaches the HRG model result from above and
starts to agree with it for $T\lsim 150$~MeV. In the
transition region this ratio is 
\begin{equation}
\frac{\chi_2^B}{\chi_2^Q} \simeq 
\fk{
(0.29-0.35)
}
\quad {\rm for}
\quad 160~{\rm MeV} \le T\le 170~{\rm MeV}\;\; ,
\label{ratioBQ}
\end{equation}
i.e., fluctuations in net electric
charge are expected to be about three to four times larger in the vicinity of
the freeze-out temperature in heavy ion collisions than net baryon
number fluctuations. It is worth noting that a comparison of  $\chi_2^B / \chi_2^Q$ 
with the ratio of proton to net charge fluctuations, which is accessible
in heavy ion collisions, will allow us to relate fluctuations in the proton
number to the fluctuations of the conserved net baryon number~\cite{Kitazawa}.

Finally, we point out that the continuum extrapolated results
presented here, and summarized in Table~\ref{tab:fluctuations} in a
temperature regime relevant to the freeze-out conditions in heavy ion
collisions, do not rely on any uncertainties in the determination of
the QCD crossover temperature or its characterization through
different observables as discussed in~\cite{hotQCDTc}. Systematic
errors on the temperature values listed in the first row of
Table~\ref{tab:fluctuations} can only come from uncertainties in the
zero-temperature observable used to determine the temperature
scale. We estimate these uncertainties to be less than $2$~MeV in our
calculation~\cite{hotQCDTc}.

\begin{table}[t]
 \begin{tabular}{|c|ccc|ccc|}
 \hline
 $T$ [MeV]& $\chi_2^B/\chi_2^{B,HRG}$ &$\chi_2^Q/\chi_2^{Q,HRG}$& $\chi_2^S/\chi_2^{S,HRG}$& $\chi_2^{BS}/\chi_2^{BS,HRG}$ &$\chi_2^{BQ}/\chi_2^{BQ,HRG}$& $\chi_2^{QS}/\chi_2^{QS,HRG}$ \\
 \hline
155 &    1.049(79)  &    0.924(36)  &    1.240(116)  &    1.353(159)  &    0.804(86)  &    1.139(74) \\
160 &    1.020(80)  &    0.895(41)  &    1.235(111)  &    1.384(135)  &    0.717(67)  &    1.144(68) \\
165 &    0.972(72)  &    0.861(44)  &    1.212(106)  &    1.356(116)  &    0.633(51)  &    1.150(67) \\
170 &    0.898(60)  &    0.818(41)  &    1.171( 96)  &    1.280( 96)  &    0.544(42)  &    1.144(64) \\
 \hline
 \end{tabular}

\begin{center}
\end{center}
\caption{
Quadratic fluctuations of net baryon number ($\delta N_B$), 
electric charge ($\delta N_Q$) and strangeness ($\delta N_S$) densities 
\fk{
and correlations among these conserved net charges
}
in the crossover region from the low to high temperature regime of QCD. 
We give results for 
quadratic fluctuations 
\fk{
and correlations
}
calculated in QCD relative to the corresponding HRG model results.
}
\label{tab:fluctuations}
\end{table}

\section*{Acknowledgments}
\addcontentsline{toc}{section}{Acknowledgements}

This work has been supported in part by contracts DE-AC02-98CH10886,
DE-AC52-07NA27344, \ DE-FC06-ER41446, DE-FG02-91ER-40628,
DE- FG02-91ER-40661, DE-KA-14-01-02 with the
U.S. Department of Energy, and NSF grants NSF10-67881,
PHY0903571, PHY08-57333, PHY07-57035,
PHY07-57333 and PHY07-03296, the Bundesministerium f\"ur Bildung und
Forschung under grant 06BI9001, the Gesellschaft f\"ur
Schwerionenforschung under grant BILAER and the Deutsche
Forschungsgemeinschaft under grant GRK881, and the EU Integrated
Infrastructure Initiative HadronPhysics2. The numerical simulations
have been performed on BlueGene/L computers at Lawrence Livermore
National Laboratory (LLNL), the New York Center for Computational
Sciences (NYCCS) at Brookhaven National Laboratory, 
US Teragrid (Texas Advanced Computing
Center), and on clusters of the USQCD collaboration in JLab and FNAL.
We thank Anton Andronic, Peter Braun-Munzinger and Krzysztof Redlich 
for discussions and for information on their current version of the 
hadron resonance gas data table.

\section*{Appendix: Choice of temperature scale and the hadron spectrum}

In this appendix we discuss the effect of setting the scale using
different observables with mass-dimension one calculated in 
zero-temperature simulations. We denote lattice observables measured in units of the
lattice spacing and calculated at a given $\beta$ by
$O_i(\beta)$. In the limit $\beta \equiv 10/g^2 \rightarrow \infty$, these approach 
their physical value $O_i^{phy}$ as
\begin{equation}
O_i(\beta) = \frac{O_i^{phy}}{\Lambda_L} 
R(\beta) \left( 1 + g^2 b_i R^2(\beta) \right) \; .
\label{Oi}
\end{equation}
where $\Lambda_L$ is the QCD scale, $R(\beta) \equiv a(\beta)
\Lambda_L$ is the $\beta$-function, and only the leading correction
${\cal O}(g^2 a^2)$, relevant to our tree-level improved staggered
formulation, has been retained.

Consider using the observable $O_i$ to define the temperature scale $T_i$. On a 
lattice of temporal size $N_\tau$, this temperature is given
in terms of $O_i^{phy}$ as
\begin{equation}
T_i(\beta) = \frac{O_i^{phy}}{O_i(\beta) N_\tau} 
\label{TOi}
\end{equation}
The ratio of any two such temperature scales is then given by 
\begin{eqnarray}
\frac{T_1(\beta)}{T_2(\beta)} &=& \frac{O_1^{phy}}{O_1(\beta)} \frac{O_2(\beta)}{O_2^{phy}} \nonumber \\
                              &=& \frac{1 + b_2 g^2 R^2(\beta)}{1 + b_1 g^2 R^2(\beta)} \nonumber \\
                              &\approx& 1 + (b_2 -b_1)g^2 R^2(\beta) \ .
\label{Trat}
\end{eqnarray}
Using Eqs.~(\ref{TOi}) and (\ref{Trat}) we can express the
observable $O_2(\beta)$ in terms of the temperature scale $T_1$ 
obtained from $O_1$ as
\begin{equation}
O_2(\beta) N_{\tau} = \frac{O_2^{phy}}{T_2} \approx
\frac{O_2^{phy}}{T_1} \left( 1 + (b_2-b_1) g^2 R^2(\beta) \right) \; .
\label{O2T}
\end{equation}
This shows that if an observable of interest ($O_2$) has a cut-off
dependence similar to that of observable ($O_1$) used to determine the
temperature scale $T_1$, {\it i.e.,} $b_2 \simeq b_1$, then $O_2(\beta)
N_\tau$, as an estimate of $O_2^{phy}/T_1$, has small cutoff effects.
Since all the $b_i - b_1$ need not be small, improving the scaling behavior of 
one observable does not, in general, imply improvement in all observables. 
 
In the low temperature regime of QCD, the relevant degrees of freedom
are hadrons with masses $M_H$. If the hadron resonance
gas is a good approximation in this regime, continuum extrapolations
of lattice data are better controlled if a temperature scale is chosen such that
all the lattice estimates of $M_H/T$ have small cut-off dependences. We find that 
using a hadronic observable such as $f_K$ improves the scaling behavior of 
the susceptibilities and correlations between charges as shown in Fig.~\ref{fig:chi2_fK_r1}. 

The pion sector is, however, different and has enhanced cut-off
effects due to taste symmetry breaking.  One does not, therefore,
expect to absorb all these effects with a choice of the temperature scale.
For this reason we had to modify the HRG analysis to compare with
lattice data for charge fluctuations which are dominated by 
contributions from the pions.



\end{document}